%% file: paper.tex
\documentclass[fleqn,a4paper,11pt]{article}
\usepackage{lineno}
\usepackage[margin=0.98in]{geometry}
\usepackage{bm}
\usepackage{amsthm}
\usepackage{amssymb}
\usepackage{latexsym}	
\usepackage{tikz}
\usetikzlibrary{quotes, backgrounds, positioning, calc, arrows.meta, decorations.pathreplacing}
\tikzset{>={Latex[color=,length=4pt,width=0pt 7]}}
\usepackage{hyperref}
\usepackage{algorithm}
\usepackage[noend]{algpseudocode}
\usepackage{url}
\usepackage{amsmath}
\usepackage{amsfonts}
\usepackage{setspace}

\newtheorem{theorem}{Theorem}
\newtheorem{lemma}[theorem]{Lemma}

\newtheorem{definition}{Definition}

\title{Sublinear algorithms for local graph centrality estimation\footnote{This is the full version of a paper accepted for publication at IEEE FOCS 2018.}}
\date{}

\author{Marco Bressan\\ Sapienza Univ.\ Roma\\bressan@di.uniroma1.it \and Enoch Peserico\\ Univ.\ Padova\\enoch@dei.unipd.it \and Luca Pretto\\Univ.\ Padova\\pretto@dei.unipd.it}

\begin{document}
\maketitle

\newcommand{\neigh}[1][\cdot]{\textsc{neigh}($#1$)}
\newcommand{\links}[1][\cdot]{\textsc{neigh}($#1$)}
\newcommand{\crawl}[1][\cdot]{\textsc{step}($#1$)}
\newcommand{\crawlNoSp}{\textsc{step}()}
\newcommand{\jump}{\textsc{jump}()}
\newcommand{\inD}{\textsc{indeg}}
\newcommand{\outD}{\textsc{outdeg}}
\newcommand{\parent}{\textsc{parent}}
\newcommand{\child}{\textsc{child}}
\newcommand{\samplenode}{\textsc{sample\-node()}}
\newcommand{\samplenodebare}{\textsc{sample\-node}}
\newcommand{\outdegree}[1]{out(#1)}
\newcommand{\indegree}[1]{in(#1)}
\newcommand{\outestim}[1]{\,\,\,\widehat{\smash[t]{\!\!\!out\!\!\!}}\,\,\,(#1)}
\newcommand{\outestimh}[1]{\,\,\,\widehat{\smash[b]{\!\!\!\smash[b]{out}\!\!\!}}\,\,\,(#1)}
\newcommand{\mindeg}[1]{outmin(#1)}
\newcommand{\maxdeg}[1]{outmax(#1)}
\algloopdefx{NoEndIf}[1]{\textbf{if} #1 \textbf{then}}
\newcommand{\mysize}{\small}
\newcommand{\parentIn}[1]{\;\scalebox{0.85}{\ensuremath{\underset{#1}\rightarrow}}\;}
\newcommand{\coef}[2]{c^{#1}_{u_{#2}}}
\newcommand{\blceil}{\Big\lceil}
\newcommand{\brceil}{\Big\rceil}
\newcommand{\E}{\mathbb{E}}
\newcommand{\cond}[3]{\mho_{#1}(#2,#3)}
\newcommand{\coefu}[2]{c_{#1}(#2)}
\newcommand{\coefc}[1]{c_{#1}}
\newcommand{\pipath}[2]{\pi_{#1}(#2)}
\newcommand{\mybeta}[2]{\beta^{#1}_{#2}}
\newcommand{\hatout}[1]{out#1}
\newcommand{\dmax}{\Delta}
\newcommand{\bareps}{\bar{\epsilon}}
\newcommand{\bfA}{\mathbf{A}}
\newcommand{\bfp}{\mathbf{p}}
\newcommand{\prob}{\text{Pr}}
\newcommand{\var}{\text{Var}}
\newcommand{\poly}{\text{poly}}
\newcommand{\polylog}{\text{polylog}}
\newcommand{\rw}{\textsc{randomwalk}}
\newcommand{\mysec}[1]{\vspace*{10pt}\noindent \textbf{#1}}

\thispagestyle{empty}
\begin{abstract}
We study the complexity of local graph centrality estimation, with the goal of approximating the centrality score of a given target node while exploring only a sublinear number of nodes/arcs of the graph and performing  a sublinear number of elementary operations.
We develop a technique, that we apply to the PageRank and Heat Kernel centralities, for building a low-variance score estimator through a local exploration of the graph.
We obtain an algorithm that, given any node in any graph of $m$ arcs,  with probability $(1-\delta)$ computes a multiplicative $(1\pm\epsilon)$-approximation of its score by examining only $\tilde{O}\big(\!\min\!\big(m^{2/3} \dmax^{1/3} d^{-2/3},\, m^{4/5} d^{-3/5} \big)\!\big)$ nodes/arcs, where $\dmax$ and $d$ are respectively the maximum and average outdegree of the graph (omitting for readability $\poly(\epsilon^{-1})$ and $\polylog(\delta^{-1})$ factors).
A similar bound holds for computational complexity.
We also prove a lower bound of $\Omega\big(\!\min\!\big(m^{1/2} \dmax^{1/2} d^{-1/2}, \, m^{2/3} d^{-1/3} \big)\!\big)$ for both query complexity and computational complexity.
Moreover, our technique yields a $\tilde{O}(n^{2/3})$ query complexity algorithm for the graph access model of Brautbar et al.~\cite{Brautbar&2010}, widely used in social network mining; we show this algorithm is optimal up to a sublogarithmic factor. These are the first algorithms yielding worst-case sublinear bounds for general directed graphs and any choice of the target node.
\end{abstract}

\clearpage

%u\end{document}

%\newpage
%\tolerance=500
%\setcounter{page}{1}
%\setstretch{0.98}
\setcounter{page}{1}
\input{introduction.tex}

\input{rel.tex}

\input{ub.tex}

\input{lowerbounds.tex}
%\input{conclusions.tex}
%\input{code.tex}

%\bibliographystyle{plainurl}
%\bibliography{biblio.bib}

%\clearpage
%\input{biblio.tex}

%\bibliographystyle{plain}
%\bibliography{general.bib}

\clearpage

\input{biblio.tex}
%\clearpage
%\input{subedges-simple.tex}

%\clearpage
%\input{subedges.tex}

%\input{subcomp.tex}

\clearpage
\input{appendix.tex}

% \renewcommand{\refname}{Additional references}
% \begin{thebibliography}{10}
% \setcounter{enumiv}{\value{firstbib}}
% \end{thebibliography}

\end{document}

%% file: introduction.tex
\section{Introduction}
\label{sec:intro}
Computing graph centralities efficiently is essential to modern network analysis.
With the advent of web and social networks, the prototypical scenario involves massive graphs on millions or even billions of nodes and arcs.
On these inputs graphs, traditional approaches such as Monte Carlo simulations and algebraic techniques are often impractical -- if not entirely useless -- since their cost can scale linearly or superlinearly with the size of the graph.
An alternative approach is that of \emph{local graph algorithms}, that, broadly speaking, work by exploring only a small portion of the graph around a given target node.
Local algorithms are justified by the fact that, often, one does not need an exact computation of the entire score vector, but only a quick approximation for a few nodes of interest.
Obviously, in exchange one hopes to drastically reduce both the running time and the portion of the graph to be fetched.
One of the best-known examples is perhaps local graph clustering~\cite{Andersen&2006,Spielman&2013,Gleich&2017}.

In this paper we address the problem of locally approximating the centrality score of a node in a graph, focusing on the PageRank and heat kernel centralities.
PageRank~\cite{BP98}  is a classic graph centrality measure with a vast number of applications including local graph clustering~\cite{Andersen&2006}, trendsetter identification~\cite{Baeza&2012}, spam filtering~\cite{Gyongyi&2004}, link prediction~\cite{Gupta&2013} and many more (see~\cite{Gleich2015} and~\cite{Chung2014}); it has been named one of the top $10$ algorithms in data mining~\cite{Wu&2007}.
Heat kernel~\cite{Chung2007} can be seen as a variant of PageRank that satisfies the heat equation.
Its applications span biological network analysis~\cite{Estrada&2005,Estrada&2010} and solving local linear systems~\cite{Chung&2015}; and, similarly to PageRank, heat kernel has a long and successful history in local graph clustering algorithms~\cite{Chung2007,Chung2009,Chung2009b,Kloster&2014,Chung&2018}.
The inputs to our problem are a directed graph $G$, a target node $v \in G$, and approximation parameters $\epsilon,\delta \in (0,1)$.
The output is a value $p(v)$ that, with probability $1-\delta$, is a multiplicative $(1\pm\epsilon)$-approximation of the centrality score $P(v)$ of $v$.
The goal is to compute $p(v)$ by fetching only a sublinear portion of $G$'s nodes and arcs, and using a sublinear number of elementary operations.
In other words, we aim at \emph{sublinear query complexity} and \emph{sublinear computational complexity}.

In the case of PageRank, the local approximation of $P(v)$ has a history dating back over a decade~\cite{CGS04,Gleich&2007,Andersen&2008,Bar-Yossef&2008a,Bar-Yossef&2008b,Bressan&2011,Borgs&2012,Borgs&2012d,Lofgren&2014b,Lofgren&2015,Lofgren&2016}.
It is well understood that PageRank can be seen as a fast-mixing random walk, which enables the approximation of all scores larger than $p$ in time $\tilde{\Theta}(1/p)$~\cite{Avrachenkov&2007,Borgs&2012,Borgs&2012d,Borgs&2013}.
However, all but $o(n)$ nodes in an $n$-node graph have score $O(1/n)$, which means the cost is $\Omega(n)$ for essentially every target node in $G$.
A complementary approach is to estimate $P(v)$ by exploring the graph backwards from $v$ towards its ancestors~\cite{CGS04,Andersen&2008,Bar-Yossef&2008a,Bar-Yossef&2008b}; however, this approach alone is subject again to a $\Omega(n)$ lower bound~\cite{Bar-Yossef&2008a,Bar-Yossef&2008b,Bressan&2011,Bressan&2013b}.
A step forward has been made by coupling the two techniques, which basically amplifies the information given by the random walks~\cite{Lofgren&2014b,Lofgren&2015,Lofgren&2016}.
The bounds obtained in this way improve on each one of the two techniques alone, but are sublinear only in expectation over $v$~\cite{Lofgren&2014b,Lofgren&2016} or only for target nodes of low degree in undirected graphs~\cite{Lofgren&2015}.
In summary, so far no sublinear bounds have been found for general directed graphs and any choice of the target node.
A similar scenario holds for the heat kernel, where research is focused on \emph{heat kernel diffusions} and their connection to Cheeger's constants and local graph clustering~\cite{Kloster&2013,Kloster&2014,Gleich&2015b,Chung&2018}, but from which no useful bound can be derived for the problem of approximating $P(v)$.

\mysec{Our Results.}
In this paper we present approximation algorithms for approximating $P(v)$ with fully sublinear worst-case query complexity and computational complexity.
Our algorithms work for general directed graphs and any choice of the target node $v$, for both PageRank and heat kernel.
(They can in principle be used for other random walk-based centralities as well, with complexity bounds depending on the choice of parameters).
For computational complexity, we use the standard RAM model.
For query complexity, we primarily use the standard graph access model of~\cite{Goldreich&1998,Goldreich&2002}, where the number of queries is essentially the number of arcs fetched.
More precisely, let $m$ be the number of arcs of $G$, and $\dmax$ and $d$ be respectively its maximum and average \emph{out}degree.
We prove:
\begin{theorem}
\label{thm:qc}
The query complexity of computing with probability $(1-\delta)$ a $(1\pm\epsilon)$-approximation of $P(v)$ is $\tilde{O}\big(\!\min\!\big( m^{2/3} \dmax^{1/3} d^{-2/3}, \, m^{4/5} d^{-3/5}\big)\big)$.
\end{theorem}
\begin{theorem}
\label{thm:cc}
The computational complexity of computing with probability $(1-\delta)$ a $(1\pm\epsilon)$-approximation of $P(v)$ is $\tilde{O}\big(\!\min\!\big(m^{3/4} \dmax^{1/4} d^{-3/4}, \, m^{6/7} d^{-5/7} \big)\big)$.
\end{theorem}
\noindent The two bounds derive from essentially the same algorithm by optimising query complexity and computational complexity \emph{separately}.
One can however keep both complexities simultaneously sublinear -- for example according to the bound of Theorem~\ref{thm:cc}, since computational complexity is an upper bound to query complexity.
In general, one can trade between the two.
Our results show one can \emph{always} break through the $\Theta(m)$ complexity barrier by polynomial factors, while all previous algorithms are no better than $O(m)$ unless one looks at special cases (e.g.\ disconnected graphs or nodes with large score).
For example, in graphs with $\Delta = O(\poly\log(n))$, which is reasonable for many social networks, our algorithms fetch only $\tilde{O}(n^{2/3})$ nodes and arcs, or perform only $\tilde{O}(n^{3/4})$ operations; approximating $P(v)$ via random walks, instead, requires $\Theta(n)$ queries and operations.
In fact our algorithms are sublinear in $n$, too, unless $m=\Theta(n^2)$.

Our second contribution are lower bounds on the query and computational complexity of approximating $P(v)$, for both PageRank and heat kernel.
Formally, we prove:
\begin{theorem}
\label{thm:lb}
$\Omega\big(\!\min\!\big(m^{1/2} \dmax^{1/2} d^{-1/2}, \, m^{2/3} d^{-1/3} \big)\big)$ queries and elementary operations are in general required to approximate $P(v)$ within a factor $O(1)$ with probability $\Omega(1)$.
\end{theorem}
\noindent 
Although weaker than the upper bounds, at the very least these lower bounds show one cannot solve the problem with e.g.\ only $\poly\log(n)$ queries and/or operations.
We also note that, unless $\Delta=\tilde{\Theta}(d)$ or $m=\tilde{\Theta}(n^2)$, our lower bounds are tighter than the $\tilde{O}(m^{1/2})$ upper bounds given by~\cite{Lofgren&2014b} for a uniform choice of $v$ (see Section~\ref{sub:rel}), proving such bounds cannot hold for \emph{every choice of} $v$.
We leave open the question of whether one can tighten our upper bounds, our lower bounds, or both.

Our final contribution are almost-tight query complexity bounds for approximating $P(v)$ under the model of~\cite{Brautbar&2010}, widely used in the field of large graph mining~\cite{Bar-Yossef&2008b,Bahmani&2012,Borgs&2012,Dasgupta&2014,Lucier&2015,Chierichetti&2016}.
The model provides a (powerful) query that returns, in one shot, all the incoming and outgoing arcs of the queried node; therefore, $n$ queries are always sufficient.
Equivalently, one can think of the query as revealing one row and one column of the adjacency matrix of $G$.
We obtain the first sublinear-query-complexity algorithm, and one that we prove optimal up to sublogarithmic factors:
\begin{theorem}
\label{thm:simple_ub}
In the model of~\cite{Brautbar&2010}, the query complexity of computing with probability $(1-\delta)$ a $(1\pm\epsilon)$-approximation of $P(v)$ is $\tilde{O}(n^{2/3})$.
Moreover, $\Omega(n^{2/3})$ queries and elementary operations are in general required to approximate $P(v)$ within a factor $O(1)$ with probability $\Omega(1)$.
\end{theorem}

\vspace*{-10pt}
\mysec{Organization of the paper.}
The rest of the introduction pins down notation and definitions.
Section~\ref{sub:rel} summarizes the state of the art.
Section~\ref{sec:upperbounds} provides a detailed walkthrough of the ideas and techniques behind our results.
All details omitted can be found in the appendix, including the pseudocode of  our algorithms (\ref{apx:pseudocode}), the adaptation for heat kernel (\ref{apx:me}), and the proofs of Theorem~\ref{thm:simple_ub} (\ref{apx:simple_ub}).

\subsection{Preliminaries and notation}
\label{sub:prelim}
We denote the directed input graph by $G=(V,A)$, and we denote by $n=|V|$ and $m=|A|$ the number of its nodes and arcs.
For simplicity we assume $n$ is known; however one can estimate it by sampling $O(\sqrt{n})$ random nodes from $G$ (see~\cite{Bressan&2015}), which leaves our bounds unchanged.
If $(u,w) \in A$ we say $u$ is a parent of $w$ and $w$ is a child of $u$, and we write $u \rightarrow w$.
We denote by $\indegree{u}$ and $\outdegree{u}$ the indegree and outdegree of $u$, and by $d = \frac{m}{n}$ and $\dmax = \max_{u \in G} \outdegree{u}$ the average and maximum outdegree of $G$.
We denote by $G[u,u',\ldots]$ the subgraph of $G$ induced by the set of nodes $\{u,u',\ldots\}$.
For simplicity we assume $G$ is free from dangling nodes ($u$ is dangling if $\outdegree{u}=0$).
This assumption makes the discussion much lighter for PageRank and can be easily lifted (see Appendix~\ref{apx:dangling}).
Unless necessary, for readability we hide in the $O()$ notation multiplicative factors depending only on the approximation parameters $\epsilon,\delta$, which in our case are (mildly) polynomial in $\epsilon^{-1}$ and polylogarithmic in $\delta^{-1}$.
Similarly, the $\tilde{O}()$ notation hides factors polylogarithmic in $n$, that in most of our bounds are actually sublogarithmic.

We denote by $\bfA$ the normalized (row-stochastic) adjacency matrix of $G$, so $a_{ij} = \frac{1}{\outdegree{i}}$ if $(i,j) \in A$ and $a_{ij} = 0$ otherwise.
The PageRank score vector $\bfp$ is then defined as:
\begin{align}
\label{eqn:pr_series}
\bfp = (1-\alpha) \sum_{\tau \ge 0} \alpha^\tau \mathbf{f}\, \bfA^\tau 
\end{align}
where $\alpha < 1$ is called the \emph{damping factor} and ensures convergence, and $\mathbf{f}$ is a stochastic \emph{preference vector}.
The heat kernel score vector is the analogous of PageRank under exponential damping:
\begin{align}
\label{eqn:hk_series}
\bfp = e^{-\alpha} \sum_{\tau \ge 0} \frac{\alpha^\tau}{\tau!} \mathbf{f}\, \bfA^\tau
\end{align}
In both cases $\lVert\bfp\rVert_1=1$, i.e.\ the scores form a probability distribution.
$P(v)$ is the entry of $\bfp$ associated to node $v$.
In our case, we set $\mathbf{f}$ to the uniform distribution $[\frac{1}{n}\ldots\frac{1}{n}]$.
Note that this implies $P(v)=\Omega(\frac{1}{n})$ for all $v$.
In most proofs we use the definitions of $P(v)$ given in Appendix~\ref{apx:definitions}; they emphasize the relationship with random walks and can be immediately derived from equations~\ref{eqn:pr_series} and~\ref{eqn:hk_series}.
Another useful equality is, for PageRank, $P(v) = \frac{1-\alpha}{n} + \sum_{u \rightarrow v} P(u) \frac{\alpha}{\outdegree{u}}$ (a slightly more involved relationship holds for heat kernel -- see Appendix~\ref{apx:me}).

For computational complexity, we adopt the standard RAM model.
For query complexity, we adopt the standard model of~\cite{Goldreich&1998,Goldreich&2002}.
Under this model, access to $G$ is provided by an oracle that answers to the following \emph{queries}: \inD($u$), that returns $\indegree{u}$; \outD($u$), that returns $\outdegree{u}$; \parent($u, i$), that returns the $i$-th parent of $u$ or $\textsc{nil}$ if $\indegree{u} < i$; \child($u, i$), that returns the $i$-th child of $u$ or $\textsc{nil}$ if $\outdegree{u} < i$.
We also allow a query \jump{}~\cite{Brautbar&2010}, that returns a node chosen uniformly at random from $G$.
Note that every call to one of these queries counts as an elementary operation, thus query complexity is a lower bound to computational complexity.
As mentioned before, we obtain bounds in the model of~\cite{Brautbar&2010}, too;
the allowed queries are \neigh[u], that returns the parents and the children of $u$, and \jump{}.

%% file: rel.tex
\section{Related work}
\label{sub:rel}
Most existing work concerns the local approximation of PageRank.
The problem itself was introduced in~\cite{CGS04}, and in its many forms has attracted considerable attention since then~\cite{Fogaras&2005,Avrachenkov&2007,Gleich&2007,Andersen&2008,Bar-Yossef&2008a,Bar-Yossef&2008b,Bressan&2011,Borgs&2012,Borgs&2012d,Lofgren&2014b,Lofgren&2015,Lofgren&2016}.

A first set of papers~\cite{Fogaras&2005,Avrachenkov&2007,Borgs&2012,Borgs&2012d} addressed the problem of sketching the scores of $G$ efficiently through random walks by using \jump{} and \child().
Since one needs $\Omega(1/P(v))$ samples in order to hit $v$, this approach is subject to a $\Omega(1/P(v))$ query and computational complexity lower bound~\cite{Borgs&2012,Borgs&2012d}, which means $\Omega(n)$ for essentially every node in $G$.
A second set of papers~\cite{Andersen&2007,Andersen&2008} focused on estimating how much each node $u \in G$ contributes to $P(v)$ (in terms of random walks, how easily one reaches $v$ from $u$).
These algorithms do not use \jump{}, and explore $G$ backwards from $v$ towards its ancestors.
%Those algorithms estimate the , and 
Although such algorithms can in principle be used to estimate $P(v)$, the lack of \jump{} makes them subject to a query complexity lower bound of $\Omega(n)$ (for Monte Carlo) or $n-o(n)$ (for Las Vegas)~\cite{Bar-Yossef&2008b,Bressan&2011,Bressan&2016}.

By combining the random walks of~\cite{Fogaras&2005,Avrachenkov&2007,Borgs&2012,Borgs&2012d} with the backward exploration of~\cite{Andersen&2007,Andersen&2008}, a set of recent papers by Lofgren et al.~\cite{Lofgren&2014b,Lofgren&2015,Banerjee&2015,Lofgren&2016} proved novel results on the local approximation of Personalized PageRank (PPR)~\cite{BP98} and related problems.
The key idea is to hit the ancestors of $v$ through random walks, which greatly reduces the necessary number of samples.
Their main result is FAST-PPR, an algorithm that approximates the PPR score of a given node within a factor $(1\pm\epsilon)$, whenever the score is larger than a given $\delta>0$.
FAST-PPR has running time $\tilde{O}(\sqrt{d/\delta})$  \textit{in expectation} over a uniform random choice of $v \in G$, which for plain PageRank means $\delta=\Theta(1/n)$ and an expected running time of $\tilde{O}(m^{1/2})$.
The framework of FAST-PPR was subsequently used to compute Markov chain multi-step transition probabilities~\cite{Banerjee&2015}, with similar average-case guarantees, and for local PageRank approximation on \emph{undirected} graphs~\cite{Lofgren&2015}, with a worst-case running time $O(n^{1/2}\,\indegree{v}^{1/2})$.
These results are encouraging, but at the same time suggest that obtaining \emph{worst-case} sublinear upper bounds for the general case is nontrivial.
Similarly to~\cite{Lofgren&2014b,Lofgren&2015,Banerjee&2015,Lofgren&2016}, in this paper we combine random walks with backward exploration.
However, we do not use the backward exploration of~\cite{Andersen&2007,Andersen&2008}, but instead we introduce two novel tools, \emph{subgraph estimators} and \emph{weighted subgraph estimators}, that make it easier to control both the variance of our estimator and the cost of its construction.
Thanks to these tools, we give the first worst-case sublinear upper bounds that hold for any directed graph $G$ and all nodes $v \in G$; incidentally, we show that the average-case upper bounds of~\cite{Lofgren&2014b} cannot hold in the worst case.

For what concerns heat kernel, existing local approximation algorithms focus on the so-called \emph{diffusions} -- essentially, the distribution of the random walk from a given seed node -- due to their relationship with local low-conductance cuts and local graph clustering~\cite{Kloster&2013,Kloster&2014,Gleich&2015b,Chung&2018}.
There exists work on efficiently computing the action of the matrix exponential on vectors~\cite{AlMohy&2011,Orecchia&2012,Gleich&2015b}, but no useful bounds can be derived for the local approximation of $P(v)$.

Finally, we shall mention recent work on the local approximation of the stationary probability of a target state $v$ in a Markov Chain~\cite{Lee&2013,Banerjee&2015,Bressan&2018}, and on the local approximation of a single entry of the solution vector of a linear system~\cite{Lee&2014,Shyamkumar&2016}.
The local approximation of $P(v)$ is a specific but nontrivial case of both, and we hope that our techniques may serve as an entry point for future developments in those directions.

%% file: ub.tex
\section{Sketch of the proofs}
\label{sec:upperbounds}
This section gives a detailed step-by-step sketch of the algorithms and proofs behind Theorem~\ref{thm:qc} and Theorem~\ref{thm:cc}.
Due to space limitations, the most technical parts have been moved to the appendix.
Here we focus on PageRank; the case of heat kernel is entirely analogous, but requires lengthy technical adaptations that can be found in Appendix~\ref{apx:me}.
The proof of Theorem~\ref{thm:simple_ub} requires adapting our algorithms as well, and can be found in Appendix~\ref{apx:simple_ub}.
The pseudocode of our algorithms is in Appendix~\ref{apx:pseudocode}.
Before proceeding, let us overview the main ideas and techniques in the order they appear in the proof sketch.
\\[0pt]
\textit{1. Random walk sampling.}
As a basic primitive we need to \emph{sample} the nodes $u \in G$ with probability equal to their score $P(u)$.
To this end we emulate the random walk, which requires $O(1)$ operations per sample in expectation.
We can then associate to each $u \in G$ an indicator random variable $\chi_u$ with expectation $P(u)$.
Estimating $P(v)$ by repeated sampling of $\chi_v$ is possible, but requires $\Omega(n)$ samples in the worst case.
\\[0pt]\textit{2. Subgraph estimators.}
Given any induced subgraph $H$ of $G$ containing $v$, by expressing $P(v)$ recursively in terms of the scores of its ancestors we define a \emph{subgraph estimator} $p_{H}(v)$ satisfying $\E[p_{H}(v)] = P(v)$.
Formally, $p_{H}(v)$ is a weighted sum of the $\chi_u$ of the nodes $u$ bordering $H$ (akin to the ``blanket sets'' of~\cite{Lofgren&2014b}).
We can take a sample of $p_{H}(v)$ using the random walk.
Unfortunately, the coefficients of the $\chi_u$ can be unbalanced and one of them can dominate, making $p_{H}(v)$ behave essentially as a single $\chi_u$, and one may still need $\Omega(n)$ samples to estimate $P(v)$.
\\[0pt]\textit{3. Weighted estimators.}
To bypass these limitations we introduce \emph{weighted estimators}, which are the weighted average of a sequence of subgraph estimators $p_{G_0}(v), p_{G_1}(v), \ldots$.
We build a weighted estimator $p_k(v)$ on the sequence of subgraphs $H=G_0,G_1,\ldots,G_k$ that we visit by exploring $G$ starting from $v$.
This requires moving from $G_{i-1}$ to $G_i$ by picking a new node $u_i$ and then \emph{expanding} it by fetching its parents and their outdegrees.
The advantage over subgraph estimators is that $p_k(v)$ exploits also the $\chi_u$ associated to the nodes \emph{inside} $G_k$, while adding a degree of freedom in weighting their coefficients.
\\[0pt]\textit{4. Building a perfect estimator.}
We prove how, by carefully choosing the nodes $u_0,\ldots,u_k$ to expand together with the weights of the subgraph estimators $p_{G_0}(v), \ldots, p_{G_k}(v)$, we can build a ``perfect'' weighted estimator $p_k(v)$ that behaves essentially as the plain sum of $k$ non-positively correlated indicator random variables $\chi_u$.
This means the variance of $p_k(v)$ is drastically lower than that of every single $p_{G_i}(v)$.
By standard concentration bounds we can then show that $\ell = \Theta(\frac{n}{k})$ samples of $p_k(v)$ suffice to estimate $P(v)$ within our approximation guarantees.
\\[0pt]\textit{5. Blacklisting heavy nodes.}
We then adapt the construction of $p_k(v)$ so to avoid expanding nodes with high indegree.
First, we take $\ell$ random walk samples; this reveals all ``heavy'' nodes having score in $\tilde{\Omega}(\ell^{-1})$, together with good estimates of their score.
The intuition is that a heavy node can have high degree.
Second, if while building $p_k(v)$ we encounter a heavy node, instead of expanding it we plug its estimate directly into $p_k(v)$.
We then show that the resulting estimator $q_k(v)$ preserves the approximation guarantees.
\\[0pt]\textit{6. Indegree inequalities.}
We then bound the total number of queries $t$ used to build $q_k(v)$ by bounding the indegree $\indegree{u_i}$ of each node we expand.
To this end we give inequalities that bound $\indegree{u_i}$ from above in terms of $m,d,\dmax$ and of $P(u_i)$.
It follows that $\indegree{u_i}$ must be small since we only expanded nodes with small $P(u_i)$.
The resulting bound is used to minimize the sum of the query complexity of all the phases (blacklisting, building, sampling).
\\[0pt]\textit{7. Approximate estimators.}
We then turn to computational complexity.
This is dominated by the construction of $q_k(v)$ and, more precisely, by the computation of the subgraph estimators $p_{G_i}(v)$.
We show that we just need an additive $\frac{\epsilon}{n}$-approximation of the coefficients of $p_{G_i}(v)$, which one can compute using $\Theta(\ln(n/\epsilon))$ sparse matrix-vector products on the normalized adjacency matrix of $G_i$.
The resulting bound is again used to minimize the sum of the computational complexity of all the phases (blacklisting, building, sampling).

\subsection{Random walk sampling}
\label{sub:rws}
Our first ingredient is a primitive to sample $v$ with probability $P(v)$.
We employ \samplenode{}, a simple routine originally introduced in~\cite{Fogaras&2005,Avrachenkov&2007} and formally defined in Appendix~\ref{apx:pseudocode}, which emulates PageRank's random walk using \jump{} and \crawl{} queries.
It is known that \samplenode{} returns node $u$ with probability $P(u)$; for convenience we give a short proof in Appendix~\ref{apx:samplenode_prob}.
Moreover, \samplenode{} has expected query and computational complexity $O(1)$, thus $\ell$ invocations have query and computational complexity $O(\ell)$ w.h.p.\ (see Appendix~\ref{apx:samplenode_queries}).
For each $u \in G$ we then let $\chi_u$ be the indicator random variable of the event that \samplenode{} returns $u$, so that $\E[\chi_u] = P(u)$ and that we can sample $\chi_u$ in expected time $O(1)$.
Clearly, the $\chi_u$ are non-positively correlated since they indicate mutually exclusive events.
The random variables $\chi_u$ will appear throughout all the rest of the paper.
Note that one could naively estimate $P(v)$ by repeatedly invoking \samplenode, but this requires $\Omega(n)$ queries since in the worst case $P(v) = O(\frac{1}{n})$.

\subsection{Subgraph estimators}
\label{sub:diffuse}
Our second ingredient is the \emph{subgraph estimator} of $P(v)$.
Recall from Section~\ref{sub:prelim} that $P(v)$ can be written recursively as $P(v) = \frac{1-\alpha}{n} + \sum_{u \rightarrow v} P(u) \frac{\alpha}{\outdegree{u}}$.
If we now replace $P(u)$ with $\chi_u$, since $\E[\chi_u] = P(u)$ we obtain a random variable $p(v)$ with expectation exactly $P(v)$.
More in general, instead of $v$ and its parents, we can consider an induced subgraph $H \subseteq G$ containing $v$, and the nodes of $G \setminus H$ having outgoing arcs that end in $H$.
Formally, define:
\begin{definition}
The \emph{frontier} of an induced subgraph $H$ of $G$ is the set of arcs $F(H)=\{(u,w) \in A : u\notin H, w\in H\}$.
We say node $u$ is \emph{on} the frontier of $H$, and we write $u \in F(H)$, if it has an outgoing arc $(u,w) \in F(H)$.
\end{definition}
\noindent 
The intuition is that $F(H)$ collects the part of $P(v)$ due to the random walks originating in $G \setminus H$.
Indeed, any random walk starting in $G \setminus H$ must go through some $u \in F(H)$ before hitting $v$.
This intuition can be formalized in the following lemma (proved in Appendix~\ref{apx:PR_conductance}):
\begin{lemma}
\label{lem:pr_diffuse}
For any induced subgraph $H$ of $G$ and any $v \in H$ it holds:
\begin{equation}
\label{eqn:PRmhox}
P(v) = \coefc{H} + \sum_{u \in F(H)} \!\!\! P(u) \cdot \coefu{H}{u}
\end{equation}
where $\coefc{H}$ and $\coefu{H}{u}$ depend only on $H$, $F(H)$ and on the outdegrees of $u \in H$ and $u \in F(H)$.
\end{lemma}
\noindent
By replacing once again $P(u)$ with $\chi_u$, we obtain:
\begin{definition}
 \label{def:diffuse}
The \emph{subgraph estimator} of $P(v)$ given by $H$ is the random variable:
\begin{align}
\label{eqn:pgv}
p_{H}(v) = \coefc{H} + \sum_{u \in F(H)} \!\! \chi_u \cdot  \coefu{H}{u}
\end{align}
\end{definition}
\noindent
By construction of $H$ we will always have $\coefc{H} \!>\! 0, \, \coefu{H}{u} \!>\! 0$.
Now, since the $\chi_u$ are non-positively correlated, we can use standard concentration bounds on $p_{H}(v)$.
The strength of the bounds depends on the coefficients $\coefu{H}{u}$: if one of them dominates, then $p_{H}(v)$ behaves essentially like a single $\chi_u$ and we may still need $\Omega(n)$ samples.
This happens for example if $H$ contains only $v$ and if all parents of $v$ have high outdegree, save one parent having outdegree $1$.
In fact there may be no $H$ making the coefficients balanced, and furthermore we know the coefficients only after fetching $H$.
Perhaps surprisingly, however, we can overcome all these limitations by simply combining subgraph estimators into a weighted sum.

\subsection{Weighted estimators}
\label{sub:weighted}
The notion of \emph{weighted estimator} of $P(v)$ is a cornerstone of our technique.
Informally, a weighted estimator is just the weighted average of a set of subgraph estimators.
Formally:
\begin{definition}
Let $G_0,\ldots,G_{k}$ be a set of induced subgraphs of $G$, each one containing $v$.
Let $\mybeta{}{0}, \ldots, \mybeta{}{k}$ be nonnegative reals such that $\sum_{i=0}^k \mybeta{}{i} = 1$.
The \emph{weighted estimator} given by $G_0,\ldots,G_{k}$ with weights $\mybeta{}{0}, \ldots, \mybeta{}{k}$ is the random variable:
\begin{align}
\label{eqn:p_m}
p_k(v) &= \sum_{i=0}^{k}\mybeta{}{i} \, p_{G_i}(v) = \sum_{i=0}^{k}\mybeta{}{i} \Big(\coefc{G_i} + \!\!\sum_{u\in F(G_i)} \!\!\!\! \chi_u \cdot \coefu{G_i}{u}\Big)
\end{align}
\end{definition}
Let us see how to employ weighted estimators.
Start by setting $u_0=v$ and $G_0 = G[u_0]$, and expand $u_0$ so to learn $F(G_0)$ and the outdegrees of all $u \in F(G_0)$.
We can then compute the coefficients of $p_{G_0}(v)$.
Now pick some $u_1 \in F(G_0)$, let $G_1 = G[u_0,u_1]$, and expand $u_1$ so to learn $F(G_1)$ and the outdegrees of all $u \in F(G_1)$.
We can then compute the coefficients of $p_{G_1}(v)$.
Now pick some $u_2 \in F(G_1)$, and so on.
We obtain a sequence of progressively larger subgraphs $G_0, \ldots, G_k$ giving subgraph estimators $p_{G_0}(v),\ldots,p_{G_k}(v)$.
By picking weights $\mybeta{}{0}, \ldots, \mybeta{}{k}$ we finally obtain our weighted estimator $p_k(v)$.

Let us take a closer look at $p_k(v)$.
For each $u$ appearing in the expression of $p_k(v)$ we let $j(u) = \min\{j : u \in F(G_j)\}$.
By rearranging the expression of Equation~\ref{eqn:p_m} so to collect together the terms pertaining a same node, we obtain:
\begin{align}
\label{eqn:p_m_2}
p_k(v) = \sum_{i = 0}^{k} \mybeta{}{i} \, \coefc{G_i} 
\;+ \;\; \sum_{i = 1}^{k} \, \chi_{u_i} \cdot \!\!\sum_{j=j(u_i)}^{i-1} \!\! \mybeta{}{j}\, \coefu{G_j}{u_i}
\;\; + \sum_{u \in F(G_{k})} \!\!\! \chi_{u} \cdot \!\!\sum_{j = j(u)}^{k} \!\! \mybeta{}{j}\, \coefu{G_j}{u}
\end{align}
\noindent
The first summation of Equation~\ref{eqn:p_m_2} gathers constant terms (containing no random variables).
The second summation gathers the nodes $u_1,\ldots,u_k$ that we have expanded and are thus in $G_k$.
The third summation gathers nodes that are still on $F(G_k)$.
Our focus is on the second summation.
We show that, by cleverly choosing $u_1, \ldots, u_k$ and $\mybeta{}{0}, \ldots, \mybeta{}{k}$, we can make the coefficients of $\chi_{u_1},\ldots,\chi_{u_k}$ \emph{all identical}, leading to (strong) concentration bounds.
This holds even though the coefficients of every single subgraph estimator $p_{G_i}(v)$ may be heavily unbalanced.

\subsection{Building a perfect weighted estimator}
\label{sub:exploring_choosing}
Let us rewrite Equation~\ref{eqn:p_m_2} more compactly.
We write the first summation, $\sum_{i = 0}^{k} \mybeta{}{i} \, \coefc{G_i}$, as $c_k$.
In the second summation, we write the coefficient $\sum_{j=j(u_i)}^{i-1} \mybeta{}{j} \coefu{G_j}{u_i}$ as $c_k(u_i)$.
In the third summation, we write the coefficient $\sum_{j=j(u)}^{k} \mybeta{}{j} \coefu{G_j}{u}$ as $c_k(u)$.
Our weighted estimator is then:
\begin{align}
\label{eqn:p_m_new}
p_k(v) = c_k + \sum_{i = 1}^{k} \,\chi_{u_i} c_k(u_i) + \!\!\!\!\sum_{u \in F(G_{k})} \!\!\!\!\! \chi_{u}\, c_k(u)
\end{align}
We call the two summations of Equation~\ref{eqn:p_m_new} respectively \emph{inner summation} and \emph{frontier summation}.
Similarly, we call their coefficients \emph{inner coefficients} and \emph{frontier coefficients}.
Ideally, we would like to make all the coefficients identical.
We settle for a slightly less ambitious target:
\begin{definition}
\label{def:balanced_estimator}
We say the weighted estimator $p_k(v)$ is \emph{perfect} if:
\begin{align*}
\begin{array}{ll}
c_k(u_i) = c_k(u_{i'}) &\text{for all } i,i' \in \{1,\ldots,k\}\\
c_k(u_i) \ge c_k(u) &\text{for all } i \in \{1,\ldots,k\} \text{ and all } u \in F(G_k)
\end{array}
\end{align*}
\end{definition}
\noindent
The heart of our algorithm consists in building a perfect weighted estimator $p_k(v)$ by exploring as little of $G$ as possible.
In fact, we show one can always build such a $p_k(v)$ by expanding exactly $k+1$ nodes.

Let us sketch the construction.
Suppose by inductive hypothesis that we have built a perfect weighted estimator $p_{k-1}(v)$ on subgraphs $G_0,\ldots,G_{k-1}$ using weights $\mybeta{}{0}, \ldots, \mybeta{}{k-1}$.
Therefore, by Definition~\ref{def:balanced_estimator} the inner coefficients $c_{k-1}(u_i)$ are all identical.
Recall the expression of $p_{k-1}(v)$ given by Equation~\ref{eqn:p_m_new} together with the definitions of $c_{k-1}(u_i)$ and $c_{k-1}(u)$.
Now note that every frontier coefficient $c_{k-1}(u)$ contains $\mybeta{}{k-1}$, while the inner coefficients $c_{k-1}(u_i)$ contain only $\mybeta{}{0},\ldots,\mybeta{}{k-2}$.
Hence, if we change $\mybeta{}{k-1}$ while rescaling $\mybeta{}{0}, \ldots, \mybeta{}{k-2}$ so that the overall sum remains $1$, we can change the ratio between inner coefficients and frontier coefficients.
In particular, we can choose $\mybeta{}{k-1}$ so that the largest (breaking ties arbitrarily) of the frontier coefficients matches the value  of the inner coefficients.
The node $u$ associated to this largest frontier coefficient is the next node $u_k$ to expand.
Indeed, we can build $p_k(v)$ from $p_{k-1}$ by moving the term concerning $u_k$ from the frontier summation to the inner summation, expanding $u_k$, and then adding to the frontier summation the new terms concerning the parents of $u_k$ that are on $F(G_k)$.
We then choose the new weight $\mybeta{}{k}=0$.
One can check that the weighted estimator $p_k(v)$ is again perfect.
Appendix~\ref{apx:pseudocode} gives the pseudocode of the construction, while Appendix~\ref{apx:construction} formally proves:
\begin{lemma}
\label{lem:building}
We can build a perfect weighted estimator $p_k(v)$ using $\sum_{i=0}^k (1+ 2\,\indegree{u_i})$ queries.
\end{lemma}
\noindent
Before bounding the query complexity and computational complexity of building $p_k(v)$, we shall analyse its concentration around $P(v)$.

\subsection{Concentration bounds for $p_k(v)$}
\label{sub:conc}
We bound the probability that $|p_k(v) - P(v)| > \epsilon P(v)$, as required by our guarantees.
Recall again the expression of $p_k(v)$ given by Equation~\ref{eqn:p_m_new}.
By Definition~\ref{def:balanced_estimator}, if $c = c_k(u_i)$ is the value of the inner coefficients, then the random variable $c^{-1}(p_k(v) - c_k)$ is a sum of non-positively correlated binary random variables with coefficients in $[0,1]$.
Furthermore, the probability that $p_k(v)$ deviates by more than a multiplicative $(1\pm\epsilon)$ from $P(v)$ is bounded by the probability that $c^{-1}(p_k(v) - c_k)$ deviates by more than a multiplicative $(1\pm\epsilon)$ from $\E[c^{-1}(p_k(v) - c_k)]$.
By the probability bounds of Appendix~\ref{apx:chernoff_bounds} we then get:
\begin{align}
\label{eqn:pml_bound_2}
\prob\big[\,|p_k(v) - P(v)| > \epsilon P(v)\big]
< 2\exp\!\Big(-\frac{\epsilon^2}{3} \E\big[c^{-1}(p_k(v) - c_k)\big]\Big)
\end{align}
Crucially, we bound from below $\E[c^{-1}(p_k(v) - c_k)]$ by restricting $p_k(v)$ to the inner summation:
\begin{align}
\E[c^{-1}(p_k(v) - c_k)] \ge \E\Big[c^{-1} \sum_{i = 1}^{k} \chi_{u_i} c_k(u_i)\Big] = \E\Big[\sum_{i = 1}^{k} \chi_{u_i}\Big]
\end{align}
Finally, since $\E[\chi_{u_i}] = P(u_i) \geq \frac{1-\alpha}{n}$ for any $u_i$, we obtain $\E[c^{-1}(p_k(v) - c_k)] \ge \frac{1-\alpha}{n}k$.
Now, if we take $\ell$ independent samples of $p_k(v)$, the expectation obviously grows to $\frac{1-\alpha}{n}k\ell$.
Therefore, if $p_k^{\ell}(v)$ is the average of $p_k(v)$ over $\ell$ samples, by Equation~\ref{eqn:pml_bound_2} we obtain:
\begin{align}
\label{eqn:conc_pkl}
\prob\big[\,|p_k^{\ell}(v) - P(v)| > \epsilon P(v)\big] < 2\exp\!\Big(\!-\frac{\epsilon^2 (1-\alpha) k \ell }{3n}\Big)
\end{align}
To get our multiplicative $(1\pm\epsilon)$-approximation of $P(v)$ with probability $1-\delta$ we must then pick $k \ell = \Theta(n \epsilon^{-2} \ln(1/\delta))$, or equivalently $\ell = \Theta(\frac{n}{k} \epsilon^{-2} \ln(1/\delta))$.

The core of our algorithm is complete: we know how to build a perfect weighted estimator $p_k(v)$ and how many samples $\ell$ to take of it.
We shall now turn to bounding the query complexity and computational complexity of building $p_k(v)$.

\subsection{Blacklisting heavy nodes}
\label{sub:blacklisting}
We now bound the query complexity $\sum_{i=0}^k (1+ 2\,\indegree{u_i})$  of building $p_k(v)$ (Lemma~\ref{lem:building}).
Unfortunately, such a complexity might be $\Theta(m)$: for instance if $G$ is a sparse graph and some $u_i$ has $\indegree{u_i} = \Theta(n)$.
We shall thus modify our algorithm so to avoid expanding nodes with high indegree or, more precisely, with high score.

First of all, we take $\ell$ samples via \samplenode.
For each $u \in G$ let $s(u)$ be the fraction of times $u$ is returned, and let $B = \{u \in G : s(u) \ge \frac{16\ln(2n/\delta)}{\epsilon^2\ell}\}$.
We call \emph{blacklisted} the nodes in $B$.
By standard arguments one can show (see Appendix~\ref{apx:vplus}):
\begin{lemma}
\label{lem:vplus}
With probability at least $1-\frac{\delta}{2}$ it holds:\\
\hspace*{10pt} \emph{1.} $\{u \in G : P(u) \ge \frac{25\ln(2n/\delta)}{\epsilon^2 \ell}\} \subseteq B$\\[0pt]
\hspace*{10pt} \emph{2.} $(1-\epsilon)P(u) \le s(u) \le (1 + \epsilon)P(u)$ for all $u \in B$
\end{lemma}
\noindent If $v \in B$, then by Lemma~\ref{lem:vplus} $s(v)$ with probability $1-\frac{\delta}{2}$ is a $(1\pm\epsilon)$-approximation of $P(v)$.
We can thus just return $s(v)$ and stop.
If instead $v \notin B$, then we build an estimator $q_k(v)$ similar to $p_k(v)$ but that avoids expanding nodes of $B$.
The idea is the following.
Suppose by inductive hypothesis that for some $k \ge 1$ we have built an estimator $q_{k-1}(v)$ which has the same form as $p_k(v)$ but is such that $u_i \notin B$ for all $i=0,\ldots,k-1$.
We then build $q_k(v)$ as in the original construction, but choosing $u_k$ among the nodes on $F(G_{k-1})$ that are not in $B$ (if there are no such nodes then we simply stop).
Therefore we set $\mybeta{}{k-1}$ so that, among all nodes on $F(G_{k-1})$ not in $B$, the largest coefficients equals the inner coefficients.
As a consequence, in $q_k(v)$ we have $u_i \notin B$ for all $i=0,\ldots,k$; while every $u \in B$ encountered during the construction is left indefinitely on the frontiers $F(G_{j(u)}), \ldots, F(G_k)$.
Formally, $q_k(v)$ has the form:
\begin{align}
\label{eqn:p_m_new_2}
q_k(v) = c_k + \sum_{i = 1}^{k} \,\chi_{u_i} c_k(u_i) + \!\!\!\!\sum_{\substack{u \in F(G_k)\\u \notin B}} \!\!\!\!\! \chi_{u}\, c_k(u) + \!\!\!\!\sum_{\substack{u \in F(G_k)\\u \in B}} \!\!\!\!\! \chi_{u}\, c_k(u)
\end{align}
Now, by construction, if restrict ourselves to the first three terms of Equation~\ref{eqn:p_m_new_2} then $q_k(v)$ is a perfect weighted estimator (Definition~\ref{def:balanced_estimator}).
We can thus apply the same bounds of $p_k(v)$ (Subsection~\ref{sub:conc}).
For the last summation, instead, we just replace each $\chi_u$ with $s(u)$ and by Lemma~\ref{lem:vplus} with probability $1-\frac{\delta}{2}$ we get a multiplicative $(1\pm\epsilon)$-approximation of the expectation of the whole sum.
Later, at sampling time, we can simply discard any sampled $u \in B$.
By a union bound, then, we get for $q_k(v)$ the same guarantees of $p_k(v)$.
We shall now conclude our query complexity bounds by bounding $\indegree{u_i}$ for $i=0,\ldots,k$.

\subsection{Indegree inequalities}
\label{sub:indegineq}
Recall that: (1) by Lemma~\ref{lem:building}, building $q_k(v)$ requires $\sum_{i=0}^k (1 + 2\,\indegree{u_i})$ queries, and (2) by Lemma~\ref{lem:vplus}, we can assume $P(u_i) < \frac{25\ln(2n/\delta)}{\epsilon^2 \ell}$ for all $i=0,\ldots,k$.
We shall then bound $\indegree{u_i}$ in terms of $P(u_i)$ and the parameters of the graph.
The intuition is that $P(u_i)$ is directly proportional to $\indegree{u_i}$, and inversely proportional to the outdegrees of $u_i$'s parents, which in turn are tied to $m, d, \dmax$.
Formally, we show (Appendix~\ref{apx:indegineq}):
\begin{lemma}
\label{lem:pc}
For any $u \in G$ it holds: \emph{1.} $\indegree{u} = O(\frac{m \dmax P(u)}{d})$, \emph{2.} $\indegree{u} = O(\frac{m  P(u)^{1/2}}{d^{1/2}})$.
\end{lemma}
\noindent
Denote $t = \sum_{i=0}^k (1 + 2 \, \indegree{u_i})$.
Now bound each term $\indegree{u_i}$ with the bounds of Lemma~\ref{lem:pc}, then in turn bound $P(u_i)$ with the bound of Lemma~\ref{lem:vplus}.
This leads to:
\begin{align}
\label{eqn:bound_t}
t = O\Big(\!\min\!\Big( \frac{k m \dmax \ln(n/\delta)}{\epsilon^2 d \ell}, \frac{k m}{\epsilon}\Big(\frac{\ln(n/\delta)}{d \ell}\Big)^{\!1/2} \Big) \! \Big)
%t = O\Big(\!\min\!\Big( \frac{k m \dmax \ln(n/\delta)}{\epsilon^2 d \ell}, \frac{k m}{\epsilon}\Big(\frac{\ln(n/\delta)}{d \ell}\Big)^{\!1/2} \Big) \! \Big)
\end{align}
\noindent
Now set $k = \frac{n \ln(1/\delta)}{\epsilon^2\ell} = \frac{m \ln(1/\delta)}{\epsilon^2 d \,\ell}$ as prescribed by the concentration bounds of Section~\ref{sub:conc}.
Then, since the blacklisting and sampling phase have query complexity $O(\ell)$, impose $\ell = t$ in order to minimize the total asymptotic query complexity $\ell+\ell+t$ of our algorithm.
We get:
\begin{align}
\label{eqn:qc_2}
\ell = O\big(\!\min\!\big( &m^{2/3} \dmax^{1/3} d^{-2/3} \ln(n/\delta)^{1/3} \ln(1/\delta)^{1/3} \epsilon^{-4/3},
\nonumber\\
&m^{4/5} d^{-3/5} \ln(n/\delta)^{1/5} \ln(1/\delta)^{2/5} \epsilon^{-6/5} \big)\!\big)
\end{align}
which becomes $\tilde{O}\big(\!\min\!\big( m^{2/3} \dmax^{1/3} d^{-2/3}, m^{4/5} d^{-3/5}\big)\big)$ if we hide factors depending only on $\epsilon$ and $\delta$.
For $\ell$ sufficiently large, we can invoke a union bound on the probability that building $q_k(v)$ requires more than $\ell$ queries (through Lemma~\ref{lem:vplus}) and on the probability of $q_k^\ell(v)$ deviating excessively from $P(v)$ (Equation~\ref{eqn:conc_pkl}), proving Theorem~\ref{thm:qc}.

\subsection{Approximate estimators}
\label{sub:approxest}
We shall finally bound the computational complexity of our algorithm.
The blacklisting and sampling phase take time $O(\ell)$.
The computationally intensive part is the construction of $q_k(v)$, or equivalently $p_k(v)$, which is dominated by the computation of the coefficients.
We can however show one just needs a good approximation of such coefficients.
Formally, we need:
\begin{definition}
$p_k'(v)$ is an \emph{additive $\bareps$-approximation} of $p_k(v)$ if $\big|\E[p_k'(v)]-\E[p_k(v)]\big| \le \bareps$.
\end{definition}
\noindent
By picking $\bareps = \Theta(\frac{\epsilon}{n})$ arbitrarily small we can then make $\E[p_k'(v)]$ arbitrarily close to $\E[p_k(v)] = P(v) \ge \frac{1-\alpha}{n}$ in a multiplicative sense.
We then build $p_k'(v)$ as we did for $p_k(v)$, so to obtain a perfect weighted estimator and keep all the concentration guarantees (with concentration around $\E[p_k'(v)]$ instead of $P(v)$).
In exchange for the additive approximation we can build $p_k'(v)$ faster than $p_k(v)$.
To this end, when computing the generic coefficients of $p_{G_i}(u)$, we ignore the contribution given by paths of length $\omega(\ln(n/\epsilon))$; we only compute the contribution of paths of length $O(\ln(n/\epsilon))$, which requires just $O(\ln(n/\epsilon))$ matrix-vector multiplications on the adjacency matrix of $G_k$.
Since this matrix has at most $t$ nonzero entries, computing the approximate coefficients of $p_{G_i}(u)$ takes time $O(t \ln(n/\epsilon))$.
Formally, we prove (Appendix~\ref{apx:pmv_approx}):
\begin{lemma}
\label{lem:pmv_approx}
An additive $\bareps$-approximation $p_k'(v)$ of $p_k(v)$ can be built in time $O(k t \ln(1/\bareps))$.
\end{lemma}
\noindent
With our choice $\bareps = \Theta(\frac{\epsilon}{n})$ this implies we can build $p_k'(v)$ in time $O(k t\ln(n/\epsilon))$.
We then use the bound $t = \tilde{O}\big(\!\min\!\big(\frac{k m \dmax }{d \, \ell}, k m (\frac{ 1 }{d\,\ell})^{1/2}\big)\big)$ given by Equation~\ref{eqn:bound_t}, and we set $\ell = \frac{m \ln(1/\delta)}{\epsilon^2 d \, k}$ for the concentration bounds.
We obtain $t = \tilde{O}\big(\!\min\!\big(\dmax k^2, k^{3/2} m^{1/2}\big)\big)$.
It follows that $q_k'(v)$ can be built in time $\tilde{O}\big(\!\min\!\big(\dmax k^3, k^{5/2} m^{1/2}\big)\big)$.
Finally, we minimize the overall computational complexity by equalling this bound with the $O(\ell) = O(\frac{m}{d\,k})$ time bound of the blacklisting and sampling phase.
We obtain:
\begin{align}
\label{eqn:cc_bound}
\ell = \tilde{O}\big(\min\!\big(m^{3/4} \dmax^{1/4} d^{-3/4}, m^{6/7} d^{-5/7} \big)\big)
\end{align}
which proves Theorem~\ref{thm:cc}.

\subsection{Remarks}
\vspace*{-10pt}
\mysec{Obliviousness to $m,d,\Delta$.}
Although we have described our algorithm as if it needed knowledge of $m, d, \dmax$, one can make it oblivious to them.
For query complexity we  proceed as follows.
Pick an initial value for $\ell$ and perform all phases on a budget of $\ell$ queries.
This is straightforward to do for blacklisting and sampling; for building $q_k(v)$, simply stop as soon as expanding $u_i$ would deplete the budget.
If at sampling time $c^{-1}(q_k^\ell(v) - c_k) = \Theta(\epsilon^{-2} \ln(1/\delta))$, by standard concentration bounds $c^{-1}(q_k^\ell(v) - c_k)$ is within $(1\pm\epsilon)$ of its expectation with probability $1-\delta$ (see Section~\ref{sub:conc}), and so $q_k^\ell(v)$ is.
Otherwise we double $\ell$ and repeat.
A similar argument holds for computational complexity.

\vspace*{-5pt}
\mysec{Portability of our techniques.}
We would spend a few words about porting our techniques to other settings.
First, we need a primitive to sample nodes with probability proportional to their score.
Second, we need $P(v)$ to be a (positive) linear combination of the scores of $v$'s ancestors.
At this point we can already build and sample a perfect estimator $p_k(v)$.
Third, we need a lower bound on each $P(u)$, or at least on $\sum_{i=1}^k P(u_i)$, so that we can attain concentration for $p_k(v)$.
Fourth, $P(u)$ must be increasing with $\indegree{u}$, so that we can blacklist nodes with large indegree.
Such ingredients are present at least partially in centralities such as Katz's~\cite{K53}, or in Markov chains.
What bounds can be obtained in these and other cases is left for future research.

%% file: lowerbounds.tex
\section{Lower Bounds}
\label{sec:lowerbounds}
We sketch the proof of Theorem~\ref{thm:lb}.
%The proof hinges on the construction of a graph $G$ of arbitrarily large size $n$, formed by two nearly identical subgraphs that one must distinguish in order to decide which of $u$ and $v$ has the higher score.
%The bound on the number of queries is obtained in expectation assuming that the decision must be made on a graph chosen uniformly at random from the family of all graphs isomorphic to $G$, and thus holds for at least one of those graphs.
For every function $d(n) \in \Omega(1) \cap O(n)$ we show a family of $n$-node graphs with average degree $d \in \Theta(d(n))$ containing a node $v$ such that, to approximate $P(v)$ within factors $O(1)$, one needs $\Omega\big(\!\min\!\big(m^{1/2} \dmax^{1/2} d^{-1/2}, \, m^{2/3} d^{-1/3} \big)\big)$ queries.
The structure of the generic graph $G$ is shown below.
%Let $d \in \Theta(d)$ be the average degree of the graph.
%Let $\Delta = n^{1/3}d^{2/3}$, and $\gamma = n^{1/3}d^{-1/3}$.
The target node $v$ has $g = n^{2/3}d^{1/3}$ parents.
One of them, $u$, has in turn $\gamma = n^{1/3}d^{-1/3}$ parents itself.
Finally, there are $n' = n-g-\gamma-2$ nodes that serve as additional children of the remaining $g-1$ parents of $v$; each parent picks as children $\Delta = n^{1/3}d^{2/3}$ of those nodes.
Note that $G$ has $m=\Theta(g\Delta) = \Theta(nd)$ arcs.
One can check that $P(v) = \Theta(\frac{\gamma}{n}) = \Theta(\frac{g}{\Delta n})$, and that by reversing the arcs between $u$ and its parents, $P(v)$ changes by a multiplicative factor $\Theta(\gamma)$.
To estimate $P(v)$ one must decide the orientation of those arcs.
%Note that we can make the change in $P(v)$ arbitrarily large by increasing $\gamma$.
It is easy to see that doing so with non-vanishing probability requires $\Theta(g)$ \parent() queries and/or $\Theta(\frac{n}{\gamma})$ \jump\ queries.
However, $\Theta(g) = \Theta(\frac{n}{\gamma}) = \Theta(m^{2/3}d^{-1/3}) = \Theta(m^{1/2} \dmax^{1/2} d^{-1/2})$.
\begin{figure*}[h!]
\begin{minipage}{.5\textwidth}
 \centering
 \input{lbgraph.tikz}
 \label{fig:graph_proof_sketch}
\end{minipage}
\begin{minipage}{.35\textwidth}
 \centering
 \input{lbgraph2.tikz}
 \label{fig:graph_proof_sketch}
\end{minipage}
\end{figure*}

%% file: lbgraph.tikz.tex
\begin{tikzpicture}[scale=0.874, every circle
node/.style={draw, thick, inner sep=0pt},
line/.style={draw, thick},
% >=Latex,
baseline={(vi.base)}]
\tikzstyle{ancestor} = [circle, draw, fill=white, minimum size=10];
\tikzstyle{strong} = [circle, draw, fill=lightgray, minimum size=8];
\tikzstyle{target} = [circle, draw, fill=white, minimum size=12, font=\large];
\tikzstyle{hidden} = [circle, draw, fill=white, minimum size=8];
\tikzset{every loop/.style={min distance=0mm, in=-110, out=-70, looseness=5}}
% \tikzstyle{hidden} = []; % This makes nodes invisibles

% ancestors horizontal relative spacing
\def\spacing{1.2};
\def\ybasecoord{0};
\def\xbasecoord{5};
\def\ancybasecoord{3};
\def\strongparent{5};
\def\spacingsink{0.4}
\def\levelzerodepth{1.9};
\def\levelzerohdepth{1.25};
\def\leveloneheight{1.7};
\def\leveltwoheight{1.6};
\def\xhorizoffset{9};
\def\nanc{6};
\def\nancminone{5};
\def\nsink{6};

\pgfmathparse{(\ybasecoord + \leveloneheight + \levelzerohdepth)/2}
\let\lzerolabelh\pgfmathresult
\pgfmathparse{(\xbasecoord + \xhorizoffset -1)}
\let\levellabelsoffset\pgfmathresult
% \def\levellabelsoffset{+12}

% \path (\xbasecoord,\lzerolabelh) ++(-5.2,0) node {\emph{level} $0 \;\, \Bigg\{$};
% \path (\xbasecoord,\leveloneheight) ++(-5.2,0) node {\emph{level} $1 \;\, \Bigg\{$};
% \path (\xbasecoord,\leveloneheight + \leveltwoheight) ++(-5.2,0) node {\emph{level} $2 \;\, \Bigg\{$};

% \path (\xbasecoord,\leveloneheight + \levelzerodepth) ++(\levellabelsoffset,0) node {\emph{level} $0$};
% \path (\xbasecoord,\leveloneheight + \levelzerohdepth) ++(\levellabelsoffset,0) node {\emph{level} $0$};
\path (\xbasecoord,\lzerolabelh) ++(\levellabelsoffset,0) node {};
\path (\xbasecoord,\leveloneheight) ++(\levellabelsoffset,0) node {};
\path (\xbasecoord,\leveloneheight + \leveltwoheight) ++(\levellabelsoffset,0) node {};

%%%%%%%%%%%%%%
%  Left side
%%%%%%%%%%%%%%

% Target node u
\node[target] (vi) at (\xbasecoord, \ybasecoord) {$v$};
%\path[->] (vi) edge[loop below] (vi);

% Parents of v
\pgfmathparse{(\nanc - 0.6)/2};
\let\hoff\pgfmathresult
\foreach \j in {0,...,\nancminone} {
  \pgfmathparse{(\j - \hoff) * \spacing};
  \let\xcoord\pgfmathresult;
  \path (vi) ++(\xcoord, \leveloneheight) node[ancestor] (w\j) {};
%   \path[->] (w\j) edge (vi);
}

% Children of parents of v
\newcount\mycount
\mycount=\strongparent
\advance\mycount by -1;
\pgfmathparse{\strongparent - 1};
\let\stpone\pgfmathresult;
\def\one{1}
\foreach \j in {0,...,\stpone} {
  \foreach \z in {1,...,\nsink} {
	\pgfmathparse{(\z - (\nsink+1)/2)*\spacingsink};
	\let\xcoords\pgfmathresult;
	\path (w\j) ++(\xcoords, \levelzerohdepth) node[hidden] (s\j\z) {};
	\path[->] (w\j) edge (s\j\z);
  }
}

% \pgfmathparse{\strongparent+1};
% \let\stptwo\pgfmathresult;
% \foreach \j in {\stptwo,...,\nancminone} {
%   \def\nsink{3};
%   \foreach \z in {1,...,\nsink} {
% 	\pgfmathparse{(\z - (\nsink+1)/2)*\spacingsink};
% 	\let\xcoords\pgfmathresult;
% 	\path (w\j) ++(\xcoords, -1) node (temp) {};
% 	\path[->] (w\j) edge[dashed] (temp);
%   }
% }

% Draw the arcs from the level-1 nodes to the target node
\pgfmathparse{(\nanc - 0.6)/2};
\let\hoff\pgfmathresult
\foreach \j in {0,...,\nancminone} {
  \pgfmathparse{(\j - \hoff) * \spacing};
  \let\xcoord\pgfmathresult;
%   \path (vi) ++(\xcoord, \leveloneheight) node[ancestor] (w\j) {};
  \path[->] (w\j) edge (vi);
}

% \path (w1) ++(-3,0) node {$1 + \left\lceil\frac{n_0 \sqrt{f(n_0)}}{\alpha}\right\rceil$};
\path (w\strongparent) ++(0,0) node {$u$};

% First parent's explicit sinks
%\def\nsink{3};
%\def\nsinkminone{2};
%\pgfmathparse{(\nsink - 0.4)/2};
%\let\hoff\pgfmathresult
%\foreach \j in {0,...,\nsinkminone} {
%  \pgfmathparse{(\j - \hoff) * \spacing};
%  \let\xcoord\pgfmathresult;
%  \path (w0) ++(\xcoord, \levelzerodepth) node[ancestor] (s\j) {};
%  \path[->] (w0) edge (s\j);
%  \path[->] (s\j) edge[loop below] (s\j);
%}

% Second-level ancestors
\def\ncloud{4};
\def\ncloudminone{3};
\pgfmathparse{(\ncloudminone)/2};
\let\hoff\pgfmathresult
\foreach \j in {0,...,\ncloudminone} {
  \pgfmathparse{(\j - \hoff)*0.4};
  \let\ycoord\pgfmathresult;
  \path (w\strongparent) ++(\spacing, \ycoord) node[hidden] (z\j) {};
  \path[->] (z\j) edge (w\strongparent);
}

\end{tikzpicture}

%% file: lbgraph2.tikz.tex
\begin{tikzpicture}[scale=0.874, every circle
node/.style={draw, thick, inner sep=0pt},
line/.style={draw, thick},
% >=Latex,
baseline={(vi.base)}]
\tikzstyle{ancestor} = [circle, draw, fill=white, minimum size=10];
\tikzstyle{strong} = [circle, draw, fill=lightgray, minimum size=8];
\tikzstyle{target} = [circle, draw, fill=white, minimum size=12, font=\large];
\tikzstyle{hidden} = [circle, draw, fill=white, minimum size=8];
\tikzset{every loop/.style={min distance=0mm, in=-110, out=-70, looseness=5}}
% \tikzstyle{hidden} = []; % This makes nodes invisibles

% ancestors horizontal relative spacing
\def\spacing{1.2};
\def\ybasecoord{0};
\def\xbasecoord{5};
\def\ancybasecoord{3};
\def\strongparent{5};
\def\spacingsink{0.4}
\def\levelzerodepth{1.9};
\def\levelzerohdepth{1.25};
\def\leveloneheight{1.7};
\def\leveltwoheight{1.6};
\def\xhorizoffset{9};
\def\nanc{6};
\def\nancminone{5};
\def\nsink{6};

\pgfmathparse{(\ybasecoord + \leveloneheight + \levelzerohdepth)/2}
\let\lzerolabelh\pgfmathresult
\pgfmathparse{(\xbasecoord + \xhorizoffset -1)}
\let\levellabelsoffset\pgfmathresult
% \def\levellabelsoffset{+12}

% \path (\xbasecoord,\lzerolabelh) ++(-5.2,0) node {\emph{level} $0 \;\, \Bigg\{$};
% \path (\xbasecoord,\leveloneheight) ++(-5.2,0) node {\emph{level} $1 \;\, \Bigg\{$};
% \path (\xbasecoord,\leveloneheight + \leveltwoheight) ++(-5.2,0) node {\emph{level} $2 \;\, \Bigg\{$};

% \path (\xbasecoord,\leveloneheight + \levelzerodepth) ++(\levellabelsoffset,0) node {\emph{level} $0$};
% \path (\xbasecoord,\leveloneheight + \levelzerohdepth) ++(\levellabelsoffset,0) node {\emph{level} $0$};
\path (\xbasecoord,\lzerolabelh) ++(\levellabelsoffset,0) node {};
\path (\xbasecoord,\leveloneheight) ++(\levellabelsoffset,0) node {};
\path (\xbasecoord,\leveloneheight + \leveltwoheight) ++(\levellabelsoffset,0) node {};

%%%%%%%%%%%%%%
%  Left side
%%%%%%%%%%%%%%

% Target node u
\node[target] (vi) at (\xbasecoord, \ybasecoord) {$v$};
%\path[->] (vi) edge[loop below] (vi);

% Parents of v
\pgfmathparse{(\nanc - 0.6)/2};
\let\hoff\pgfmathresult
\foreach \j in {0,...,\nancminone} {
  \pgfmathparse{(\j - \hoff) * \spacing};
  \let\xcoord\pgfmathresult;
  \path (vi) ++(\xcoord, \leveloneheight) node[ancestor] (w\j) {};
%   \path[->] (w\j) edge (vi);
}

% Children of parents of v
\newcount\mycount
\mycount=\strongparent
\advance\mycount by -1;
\pgfmathparse{\strongparent - 1};
\let\stpone\pgfmathresult;
\def\one{1}
\foreach \j in {0,...,\stpone} {
  \foreach \z in {1,...,\nsink} {
	\pgfmathparse{(\z - (\nsink+1)/2)*\spacingsink};
	\let\xcoords\pgfmathresult;
	\path (w\j) ++(\xcoords, \levelzerohdepth) node[hidden] (s\j\z) {};
	\path[->] (w\j) edge (s\j\z);
  }
}

% \pgfmathparse{\strongparent+1};
% \let\stptwo\pgfmathresult;
% \foreach \j in {\stptwo,...,\nancminone} {
%   \def\nsink{3};
%   \foreach \z in {1,...,\nsink} {
% 	\pgfmathparse{(\z - (\nsink+1)/2)*\spacingsink};
% 	\let\xcoords\pgfmathresult;
% 	\path (w\j) ++(\xcoords, -1) node (temp) {};
% 	\path[->] (w\j) edge[dashed] (temp);
%   }
% }

% Draw the arcs from the level-1 nodes to the target node
\pgfmathparse{(\nanc - 0.6)/2};
\let\hoff\pgfmathresult
\foreach \j in {0,...,\nancminone} {
  \pgfmathparse{(\j - \hoff) * \spacing};
  \let\xcoord\pgfmathresult;
%   \path (vi) ++(\xcoord, \leveloneheight) node[ancestor] (w\j) {};
  \path[->] (w\j) edge (vi);
}

% \path (w1) ++(-3,0) node {$1 + \left\lceil\frac{n_0 \sqrt{f(n_0)}}{\alpha}\right\rceil$};
\path (w\strongparent) ++(0,0) node {$u$};

% First parent's explicit sinks
%\def\nsink{3};
%\def\nsinkminone{2};
%\pgfmathparse{(\nsink - 0.4)/2};
%\let\hoff\pgfmathresult
%\foreach \j in {0,...,\nsinkminone} {
%  \pgfmathparse{(\j - \hoff) * \spacing};
%  \let\xcoord\pgfmathresult;
%  \path (w0) ++(\xcoord, \levelzerodepth) node[ancestor] (s\j) {};
%  \path[->] (w0) edge (s\j);
%  \path[->] (s\j) edge[loop below] (s\j);
%}

% Second-level ancestors
\def\ncloud{4};
\def\ncloudminone{3};
\pgfmathparse{(\ncloudminone)/2};
\let\hoff\pgfmathresult
\foreach \j in {0,...,\ncloudminone} {
  \pgfmathparse{(\j - \hoff)*0.4};
  \let\ycoord\pgfmathresult;
  \path (w\strongparent) ++(\spacing, \ycoord) node[hidden] (z\j) {};
  \path[->] (w\strongparent) edge (z\j);
}

\end{tikzpicture}

%% file: appendix.tex
\appendix
\section{TECHNICAL DETAILS}
\label{sec:appendix}

\subsection{Pseudocode}
\label{apx:pseudocode}
The routine ApproxCentrality$(v,\epsilon,\delta)$ is the entry point of our whole algorithm.
All other routines are invoked during the execution.
As noted in Section~\ref{sec:upperbounds}, although for simplicity we write the algorithm as if it knew the parameters $\Delta,d$ of the graph, it can easily made oblivious to them by progressively increasing the query budget.

\renewcommand{\thealgorithm}{}
\begin{algorithm}
\caption{ApproxCentrality$(v,\epsilon,\delta)$}
\begin{algorithmic}[1]
\State $\ell, k \leftarrow$ from the bounds of Section~\ref{sec:upperbounds}
\State $B, s[\,] \leftarrow$ BlackList($\epsilon$, $\delta$, $\ell$)
\State $c_k, c_k[\,] \leftarrow$ BuildEstimator($v$, $k$, $B$)
\State \Return SampleEstimator($c_k$, $c_k[\,]$, $\ell$, $B$) $+ \sum_{u \in B} c_k[u]s[u]$
\end{algorithmic}
\end{algorithm}

\begin{algorithm}[H]
\caption{Blacklist($\epsilon$, $\delta$, $\ell$)}
\begin{algorithmic}[1]
\State $s[\,] \leftarrow$ empty dictionary with default value $0$
\For{$i=1,\ldots,\ell$}
\State $u \leftarrow$ \samplenode{}
\State $s[u] \leftarrow s[u] + \frac{1}{\ell}$
\If{$s[u] \ge \frac{16\ln(2n/\delta)}{\epsilon^2\ell}$}
  \State $B \leftarrow B \cup u$
\EndIf 
\EndFor
\State \Return $B, s[\,]$
\end{algorithmic}
\end{algorithm}

\begin{algorithm}[H]
\caption{BuildEstimator($v$, $k$, $B$)}
%\Comment a simpler version of the original algorithm, starting with the target node instead of one of its parents
\begin{algorithmic}[1]
\State $u_0 \leftarrow v$
\State expand $u_0$ and set $H \leftarrow G[u_0]$
\State $c \leftarrow c_{H}(u_0)$ \Comment inner coefficient
\State $c_0 \leftarrow c_{H}$ \Comment subgraph coefficient
\State $c_0[\,]\leftarrow$ empty dictionary with default value $0$
\For{$u \in F(H)$} 
  \State $c_0[u] \leftarrow c_{H}(u)$ \Comment frontier coefficients
\EndFor
\For{$i=1,\ldots,k$}
  \State $\Delta[\,]\leftarrow$ empty dictionary
  \For{$u \in F(H) \setminus B$}
    \State $\Delta[u] \leftarrow c - c_{i-1}[u]$
  \EndFor
  \State $c_H, c_H[\,] \leftarrow$ ComputeCoefs$(H,\epsilon/n)$
  \State $u_i \leftarrow \arg \min\{\Delta[u]/(\Delta[u] + c_{H}[u]) \,:\, u \in F(H) \setminus B \}$
  \State $\beta \leftarrow \Delta[u_i]/(\Delta[u_i] + c_{H}[u_i])$
%  \State $\beta_i^i \leftarrow \min_{u\in\text{F}}\big\{ \Delta_u/(\Delta_u + \frac{c_{G_{i}}(u)}{\text{\hatout{[u]}}})\big\}$
%  \State $u_i \leftarrow \arg \min_{u\in\text{F}} \Delta_i(u)/(\frac{c_{G_{i-1}}(u)}{degs(u)}+\Delta_i(u))$
%  \State $\beta_i^i \leftarrow \Delta_i(u_i)/(\frac{c_{G_{i-1}}(u_i)}{degs(u_i)}+\Delta_i(u_i))$
  \State $c \leftarrow (1-\beta) c$
  \State $c_i \leftarrow (1-\beta) c_{i-1} + \beta\, c_{H}$
%  \State $c^{\text{in}} \leftarrow (1-\beta_i^i) c^{\text{in}} $ \Comment $c^{\text{in}}_i$
  \State $c_i[\,]\leftarrow$ empty dictionary with default value $0$
  \For{$u \in F(H)$}
%       \State $c_{G_i}(u) \leftarrow $ compute
    \State $c_i[u] \leftarrow (1-\beta) c_{i-1}[u] + \beta\, c_{H}[u]$ 
  \EndFor
  \State expand $u_i$ and set $H \leftarrow G[u_0,\ldots,u_i]$
\EndFor
\State \Return $c_k$,\, $c_k[u_1,\ldots,u_k] \cup c_k[u : u \in F(H)]$)
\end{algorithmic}
\end{algorithm}

\begin{algorithm}[H]
\caption{SampleEstimator($c_k$, $c_k[\,]$, $\ell$, $B$)}
\begin{algorithmic}[1]
\State $p \leftarrow 0$
\For{$i=1,\ldots,\ell$}
\State $u \leftarrow$ \samplenode{}
%\State $(\mathcal{P},\mathcal{C}) \leftarrow$ \neigh[u]
%\State $\outdegree{u} \leftarrow |\mathcal{C}|$
\If{$p \notin B$}
\State $p \leftarrow p \,+\, \ell^{-1}(c_k \, + \, c_k[u])$
\EndIf
\EndFor
\State \Return $p$
\end{algorithmic}
\end{algorithm}

\renewcommand{\thealgorithm}{}
\begin{algorithm}[H]
\label{alg:photo}\caption{SampleNode$()$}
\begin{algorithmic}[1]
\State $u \leftarrow$ \jump
\Loop
\State with probability $(1-\alpha)$ \textbf{return} $u$
\State $\outdegree{u} \leftarrow $ \outD[u]
\If{$\outdegree{u} == 0$}
\State $u \leftarrow$ \jump
\Else
\State $u \leftarrow$ \child($u$, randint($\outdegree{u}$))
\EndIf
\EndLoop
\end{algorithmic}
\end{algorithm}

\renewcommand{\thealgorithm}{}
\begin{algorithm}[H]
\label{alg:coefs}\caption{ComputeCoefs$(H, \bar{\epsilon})$}
\begin{algorithmic}[1]
\State $k \leftarrow |V_H|$
\State $\bfA_H \leftarrow \bfA_G$ restricted to $H$
\State $\mathbf{r}, \mathbf{c} \leftarrow [0,\ldots,0]$ of length $k$
\State $\mathbf{r}[v], \mathbf{c}[v] \leftarrow 1$
\For{$i = 1,\ldots,\ln(1/\bar{\epsilon})$}
  \State $\mathbf{r} \leftarrow \alpha\, \bfA_H\, \mathbf{r} $
  \State $\mathbf{c} \leftarrow \mathbf{c} + \mathbf{r}$
\EndFor
\State $c_H \leftarrow 0,\; c_{H}[\,] \leftarrow $ empty dictionary with default value $0$
\For{$z \in H$}
\State $c_{H} \leftarrow c_{H} + \frac{1-\alpha}{n} \mathbf{r}[z]$
\EndFor
\For{$(u,w) \in F(H)$}
\State $c_{H}[u] \leftarrow c_{H}[u] + \frac{\alpha}{\outdegree{u}} \mathbf{r}[w]$
\EndFor
\State\Return $c_H, c_H[\,]$
\end{algorithmic}
\end{algorithm}

\subsection{Probability bounds}
\label{apx:chernoff_bounds}
We give Chernoff-type probability bounds that are repeatedly used in our analysis.
These bounds can be found in e.g.~\cite{Auger&2011} and can be derived from~\cite{Panconesi&1997}.
Let $X_1,\ldots,X_n$ be binary random variables. We say that $X_1,\ldots,X_n$ are non-positively correlated if for all $I \subseteq \{1,\ldots,n\}$ we have:
\begin{align}
\prob[\forall i \in I: X_i=0] &\leq \prod_{i \in I} \prob[X_i=0] \\
\prob[\forall i \in I: X_i=1] &\leq \prod_{i \in I} \prob[X_i=1]
\end{align}
The following lemma holds:
\begin{lemma}
\label{lem:chernoff}
Let $X_1,\ldots,X_n$ be independent or, more generally, non-positively correlated binary random variables. Let $a_1,\ldots,a_n \in [0,1]$ and $X=\sum_{i=1}^{n}a_iX_i$. Then, for any $\epsilon > 0$, we have:
\begin{align}
\prob[X < (1-\epsilon)\E[X]] &< e^{-\frac{\epsilon^2}{2}\E[X]} \\
\prob[X > (1+\epsilon)\E[X]] &< e^{-\frac{\epsilon^2}{2+\epsilon}\E[X]} 
\end{align}
\end{lemma}
\noindent
Note that Lemma~\ref{lem:chernoff} applies if $X_1,\ldots,X_n$ are indicator variables of mutually disjoint events, or can be partitioned into independent families $\{X_1,\ldots,X_{i_1}\}$, $\{X_{i_1+1},\ldots,X_{i_2}\}$, \ldots of such variables.

\subsection{Additional definitions}
\label{apx:definitions}
A \emph{path} $\pi$ from $z$ to $v$ is a sequence of arcs of $G$ such that either $\pi = \emptyset$ and $z = v$, or $\pi$ is the concatenation of a path $\pi'$ from $z$ to some $w$ such that $(w,v) \in G$ with the arc $(w, v)$.
The \emph{length} $|\pi|$ of a path $\pi$ is the number of arcs in it.
%Also, we denote by $\outdegree{w}$ the outdegree of node $w$.
The \emph{weight} of a path $\pi$ is:
\begin{align}
\rho_{\pi} = \prod_{(x,y)\in \pi} \frac{1}{\outdegree{x}}
\end{align}
The \emph{resistance} $\mho^{\pi}$ of a path $\pi$ is $\alpha^{|\pi|} \rho_\pi$.
%\begin{align}
%\mho^{\pi} = \prod_{(x,y)\in \pi} \frac{\alpha}{\outdegree{x}}
%= \alpha^{|\pi|} \prod_{(x,y)\in \pi}\frac{1}{\outdegree{x}}
%\end{align}
For PageRank, $\mho^{\pi}$ is the probability that the PageRank random walk, starting on the first node of $\pi$, follows exactly $\pi$.
Given a subgraph $H \subseteq G$, we say $\pi \in H$ if each arc of $\pi$ is in $H$.
We denote by $\Pi_{H}(z,v)$ be the set of all paths from $z$ to $v$ in $H$.
The \emph{resistance} of $H$ from $z$ to $v$ is the sum of the resistances of all paths in $\Pi_{H}(z,v)$:
\begin{align}
\cond{H}{z}{v} = \sum_{\pi \in \Pi_{H}(z,v)} \!\!\!\!\mho^{\pi}
\end{align}
Note that $\cond{H}{z}{v}$ is always finite.
Note also that $\cond{H}{v}{v}$ contains the empty path from $v$ to itself, with resistance $1$.

Finally, from equations~\ref{eqn:pr_series} and~\ref{eqn:hk_series} (Section~\ref{sub:prelim}) one can derive the standard equalities:
\begin{align}
\label{eqn:pr_def}
\text{PageRank:} \quad  P(v) = \frac{\eta}{n} \sum_{\tau \ge 0} \, \sum_{z \in G} \, \sum_{\substack{\pi \in \Pi_{G}(z,v) \\ |\pi| = \tau }} \!\!\! \alpha^{\tau} \rho_{\pi}, \quad\eta = \big(\sum_{\tau\ge 0} \alpha^{\tau}\big)^{-1} = 1-\alpha
\\
\label{eqn:me_def}
\text{heat kernel:} \quad P(v) = \frac{\eta}{n} \sum_{\tau \ge 0} \, \sum_{z \in G} \, \sum_{\substack{\pi \in \Pi_{G}(z,v) \\ |\pi| = \tau }} \!\!\!  \frac{\alpha^{\tau} \rho_{\pi}}{\tau!}, \quad\eta = \big(\sum_{\tau\ge 0} \frac{\alpha^{\tau}}{\tau!}\big)^{-1} = e^{-\alpha}
\end{align}

\subsection{Proof of Lemma~\ref{lem:pr_diffuse}}
\label{apx:PR_conductance}
\label{apx:conductance}
From Equation~\ref{eqn:pr_def} and the definition of path resistance $\mho^{\pi}$ we obtain:
\begin{align}
P(v)
%\; = \; \sum_{z \in G} \frac{1-\alpha}{n} \sum_{\pi \in \Pi_{G}(z,v)} \prod_{(x,y)\in \pi} \frac{\alpha}{\outdegree{x}}
\; = \; \sum_{z \in G} \, \sum_{\pi \in \Pi_{G}(z,v)} \frac{1-\alpha}{n} \, \mho^{\pi}
\end{align}
Now, by considering separately the paths $\pi \in H$ and the paths $\pi \notin H$:
\begin{align}
\label{eqn:Pv_mho}
P(v) \; = \; \sum_{z \in H} \, \sum_{\substack{\pi \in \Pi_{H}(z,v)}} \!\! \frac{1-\alpha}{n} \, \mho^{\pi}
\;+\; \sum_{z \in G} \, \sum_{\substack{\pi \in \Pi_{G}(z,v) \\ \pi \notin H}} \!\!\frac{1-\alpha}{n} \, \mho^{\pi}
\end{align}
The first term equals $\sum_{z \in H} \frac{1-\alpha}{n} \, \cond{H}{z}{v}$.
For the second term, observe that any $\pi \notin H$ from $z$ to $v$ can be uniquely decomposed into a head path $\pi'$ from $z$ to some $u \in F(H)$, an arc $(u,w) \in F(H)$, and a tail path $\pi'' \in H$ from $w$ to $v$.
We can then split each such $\pi \notin H$ on the arc $(u,w)$ and gather terms using the same arc $(u,w)$.
Moreover, since the probability of following $(u,w)$ once in $u$ is $\frac{\alpha}{\outdegree{u}}$, the second term becomes:
\begin{align}
\sum_{z \in G} \, \sum_{\pi \in \Pi_{G}(z,v)} \!\!\! \frac{1-\alpha}{n} \, \mho^{\pi}
&\;= \sum_{(u,w) \in F(H)} \Big( \sum_{z \in G} \sum_{\pi' \in \Pi_{G}(z,u)} \frac{1-\alpha}{n} \, \mho^{\pi'}  \Big) \frac{\alpha}{\outdegree{u}} \, \sum_{\substack{\pi'' \in \Pi_{H}(w,v)}} \!\!\!\! \mho^{\pi''}
\end{align}
The quantity between brackets equals $P(u)$, while the last summation equals $\cond{H}{w}{v}$.
Thus the expression equals $\sum_{(u,w) \in F(H)} P(u) \frac{\alpha}{\outdegree{u}} \cond{H}{w}{v}$.

We then define:
\begin{align}
c_H &= \sum_{z \in H} \frac{1-\alpha}{n} \, \cond{H}{z}{v}\\
\coefu{H}{u} &= \sum_{w:(u,w)\in F(H)} \frac{\alpha}{\outdegree{u}}\,\cond{H}{w}{v}
\end{align}
It follows from Equation~\ref{eqn:Pv_mho} that $P(v) = c_H + \sum_{u\in F(H)} P(u)  \cdot \coefu{H}{u}$ and the proof is over.

\subsection{Output of \samplenode{}}
\label{apx:samplenode_prob}
We prove:
\begin{lemma} 
\label{lem:samplenode}
\samplenode{} returns node $u$ with probability $P(u)$.
\end{lemma}
\begin{proof}
From Equation~\ref{eqn:pr_series} one can check that the following procedure yields $u$ with probability $P(u)$.
First, pick $L \ge 0$ from the distribution given by $\prob[L=\tau] = (1-\alpha)\alpha^\tau$.
Then, pick $z$ uniformly at random in $G$.
Finally, walk $L$ steps from $z$, at each step moving to a children of the current node chosen uniformly at random.
%From the equivalence $P(v) = \frac{1-\alpha}{n} + \sum_{u \rightarrow v} P(u) \frac{\alpha}{\outdegree{u}}$ (see Subsection~\ref{sub:prelim}) one can verify the following standard fact.
%$P(v)$
%It is equivalent to the following rule, to be applied at each step.
%With probability $(1-\alpha)$, we jump to a node chosen uniformly at random in $G$.
%With probability $\alpha$, we move to a children of the current node chosen uniformly at random; but if the current node is dangling, we imagine it has $n$ virtual arcs to all nodes of $G$.
%Now the stationary probability of the latest jump being $\tau$ steps in the past, i.e.\ of having followed after the jump exactly $\tau$ arcs, whether virtual or not, is $(1-\alpha)\alpha^\tau$.
%Therefore if we perform a random jump and then follow a path on $\tau$ arcs (virtual or not) with probability exactly $(1-\alpha)\alpha^\tau$, then the ending node will be $v$ with probability $P(v)$.
This is precisely what \samplenode{} does, with the sole difference of drawing $L$ during the execution.
Finally, note that \samplenode{} is designed so to jump to a node uniformly chosen at random in $G$ in case the current node is dangling.
\end{proof}

\subsection{Complexity of SampleNode()}
\label{apx:samplenode_queries}
\begin{lemma} 
\label{lem:samplenode_queries}
The probability that $\frac{2\ell}{(1-\epsilon)(1-\alpha)}$ queries are not sufficient to complete $\ell$ calls to \samplenode{} is less than $e^{-\ell\epsilon^2/2(1-\epsilon)}$.
\end{lemma}
\begin{proof}
Each time line 3 is executed, with probability $(1-\alpha)$ \samplenode{} terminates.
Now, if a sequence of \samplenode{} calls uses $q$ queries, then line 3 is executed at least $q/2$ times, and at each execution \samplenode{} terminates with probability $(1-\alpha)$.
The expected number of \samplenode{} calls one can complete with $q$ queries is thus at least $q(1-\alpha)/2$, and by the bounds of Appendix~\ref{apx:chernoff_bounds}, the probability that the number of completed calls does not reach $(1-\epsilon)q(1-\alpha)/2$ is less than $e^{-q(1-\alpha)\epsilon^2/4}$.
Setting $q = \frac{2\ell}{(1-\epsilon)(1-\alpha)}$ gives the thesis.
\end{proof}

%\subsection{Proof of Lemma~\ref{lem:pml_conc}}
%\label{apx:pml_conc}
%Consider the random variable:
%\begin{align}
%\label{eqn:pm_core}
%\frac{\ell}{c}(p_m^\ell(v) - c_m) = \sum_{h=1}^\ell \Big( \sum_{i = 1}^{m} \chi_{u_i}^h \frac{c_m(u_i)}{c} +\!\! \sum_{u \in F(G_{m})}\!\!\!\!\! \chi_{u}^h \frac{c_m(u)}{c} \Big)
%\end{align}
%where $c$ is the maximum among all inner and frontier coefficients of $p_m(v)$.
%Clearly, $p_m^\ell(v)$ is the sum of the nonnegative constant $c_m$ and of $\frac{c}{\ell}$ times $\frac{\ell}{c}(p_m^\ell(v) - c_m)$, and thus deviates by more than a factor $1 \pm \epsilon$ from $P(v)$ only if $\frac{\ell}{c}(p_m^\ell(v) - c_m)$ deviates by more than a factor $(1 \pm \epsilon)$ from its own expectation.
%Since $\frac{\ell}{c}(p_m^\ell(v) - c_m)$ is by construction a sum of non-positively correlated indicator random variables with weights in $[0,1]$, the probability of the latter event can be bounded by applying the probability bounds of Appendix~\ref{apx:chernoff_bounds}, which yield for any $\epsilon$ in $[0,1]$:
%\begin{align}
%\prob\big[|p_m^\ell(v) - P(v)| > \epsilon P(v)\big]
%\leq 2e^{-\epsilon^2 \E[\frac{\ell}{c}(p_m^\ell(v) - c_m)]/3}
%\end{align}

\subsection{Proof of Lemma~\ref{lem:building}}
\label{apx:construction}
%We prove a slightly stronger result anticipated in the sketch and needed by the proof: if the outdegree of every node in $G$ is known beforehand, then with $m+1$ \neigh{} queries one can choose $u_1, \ldots, u_m$ and $\mybeta{m}{0},\mybeta{m}{1},\ldots,\mybeta{m}{m}$ so that $p_m(v)$ is $1$-balanced and there exists a frontier node whose coefficient matches the inner coefficients.
We prove the claim by induction.
Recall that we want the coefficients of $p_k(v)$ to satisfy:
\begin{align}
\label{eqn:balanced_coefficients_inner}
&\forall i,i' : 1 \leq i,i'\leq k: \;
c_k(u_i) = c_k(u_{i'})
\\\label{eqn:balanced_coefficients_outer}
&\forall u \in F(G_k): \;
c_k(u) \leq c_k(u_k)
%\\\label{eqn:balanced_coefficients_next}
%\text{3. }&\text{if } F(G_m) \neq \emptyset  \text{, then }\nonumber\\
%&\exists u \in F(G_m): \; c_m(u) = c_m(u_m)
\end{align}
We start with $k=1$.
Note that if $F(G_0) = \emptyset$ or $F(G_1) = \emptyset$ then we can stop and compute $P(v)$ exactly.
Let $\mybeta{}{0}=1$ and $\mybeta{}{1}=1-\mybeta{}{0}=0$.
Let $u_0=v$, and let $u_1$ be the parent of $v$ of smallest outdegree (breaking ties arbitrarily).
%Therefore $G_0=G[v]$ and $G_1=G[u_1,v]$.
 % (for $k=0$ the inner summation of $p_k(v)$ is empty).
The inner summation of $p_1(v)$ has only one term, so Equation~\ref{eqn:balanced_coefficients_inner} is trivially satisfied.
We thus focus on Equation~\ref{eqn:balanced_coefficients_outer}.
%Note that all $u \in F(G_0)$ are parents of $v$.
%Let $u_1$ be the one with smallest outdegree, breaking ties arbitrarily.
%Set $\mybeta{}{0} = 1$ and $\mybeta{}{1}=1-\mybeta{}{0}=0$.
By the definitions of $c_k(u_i)$ and of $\coefu{G_0}{u_i}$ (Appendix~\ref{apx:PR_conductance}), we have:
\begin{align}
c_1(u_1) &= \mybeta{}{0} \, \coefu{G_0}{u_1} = \frac{\alpha \, \cond{G_0}{v}{v}}{\outdegree{u_1}} 
\label{eqn:c1_unrolled}
\end{align}
Similarly, by recalling the definition of $c_k(u)$, for each $u \in F(G_{1})$ we have:
\begin{align}
c_1(u) &= \sum_{j=j(u)}^{0} \!\! (1-\mybeta{}{1}) \coefu{G_0}{u} \,+\,  \mybeta{}{1} \coefu{G_1}{u}
= \sum_{j=j(u)}^{0} \!\! \frac{\alpha \, \cond{G_0}{v}{v}}{\outdegree{u}}
\label{eqn:c1_unrolled_2}
\end{align}
If $j(u)=1$ then the rightmost summation is empty, so $c_1(u) = 0 \le c_1(u_1)$.
If instead $j(u)=0$ then $u$ is a parent of $v$, and thus  $c_1(u) = \frac{\alpha \, \cond{G_0}{v}{v}}{\outdegree{u}} \le \frac{\alpha \, \cond{G_0}{v}{v}}{\outdegree{u_1}} \le c_1(u_1)$ by our choice of $u_1$.

Suppose now by inductive hypothesis $p_{k-1}(v)$ for $k-1 \ge 1$ is a perfect weighted estimator, so it satisfies Equations~\ref{eqn:balanced_coefficients_inner} and \ref{eqn:balanced_coefficients_outer}.
%Let $u_0=v,u_1, \ldots, u_{k-1}$ and $\mybeta{}{0},$ $\mybeta{}{1},$ $\ldots,\mybeta{}{k-1}$ be the associated Therefore  and~\ref{eqn:balanced_coefficients_next} hold for $p_{k-1}(v)$.
Assume once again $\mybeta{}{k-1}=0$, and assume $F(G_{k-1}) \ne \emptyset$, or we could stop and compute $P(v)$ exactly.
%Consider the sets $u_1, \ldots, u_{m}$ and $\mybeta{m}{0},\mybeta{m}{1},\ldots,\mybeta{m}{m}$ defined as follows: $u_{m}$ is some node $u'$ satisfying Equation~\ref{eqn:balanced_coefficients_next}, while $\mybeta{m}{i} = (1-\mybeta{m}{m})\mybeta{m-1}{i}$ for all $i = 1, \ldots, m-1$, for some $\mybeta{m}{m}$ to be chosen later.
Now suppose for some $\beta > 0$ we set:
\begin{align}
p_k(v) = (1-\beta)\,p_{k-1}(v) + \beta \, p_{G_{k-1}}(v)
\end{align}
so that, by replacing the expressions of $p_{k-1}(v)$ and $p_{G_{k-1}}(v)$:
\begin{align}
\label{eqn:pkv_next}
p_k(v) 
%=&\ (1-\beta)c_{k-1} 
%+ \sum_{i = 1}^{k-1} \, \chi_{u_i} \cdot (1-\beta)c_{k-1}(u_i)
%+ \!\sum_{u \in F(G_{k-1})} \!\!\!\! \chi_{u} \cdot (1-\beta)c_{k-1}(u)
%\nonumber\\
%&+ \beta c_{G_{k-1}}
%+ \!\sum_{u \in F(G_{k-1})} \!\!\!\!\!\! \chi_{u} \cdot \beta\, c_{G_{k-1}}(u)\\
=&\ (1-\beta)c_{k-1} + \beta c_{G_{k-1}}
+ \sum_{i = 1}^{k-1} \, \chi_{u_i} \cdot (1-\beta)c_{k-1}(u_i)
\nonumber\\&+ \!\sum_{u \in F(G_{k-1})} \!\!\!\!\!\! \chi_{u} \cdot \big((1-\beta)c_{k-1}(u) + \beta\, c_{G_{k-1}}(u)\big)
\end{align}
Clearly, for $\beta=0$ we have $p_k(v) = p_{k-1}(v)$.
Now, if we progressively increase $\beta$, all inner coefficients will shrink together by the same factor $(1-\beta)$, and in the frontier coefficient of $u$ the term $\beta \, c_{G_{k-1}}(u)$ will grow.
At some point the largest frontier coefficient (breaking ties arbitrarily) will therefore match the inner coefficients.
Formally, for each $u \in F(G_{k-1})$ let then $\Delta(u) = c_{\text{inner}} - c_{k-1}(u)$, where $c_{\text{inner}} = c_{k-1}(u_i)$.
Note that by hypothesis $\Delta(u) \ge 0$.
Now let:
\begin{align}
\label{eqn:beta_oneshot}
\bar u &= \arg \min_{u \in F(G_{k-1})} \frac{\Delta(u)}{\Delta(u) + c_{G_{k-1}}(u)}, \qquad \beta = \frac{\Delta(\bar u)}{\Delta(\bar u) + c_{G_{k-1}}(\bar u)}
\end{align}
It is easy to check that $(1-\beta)c_{\text{inner}} = (1-\beta) c_{k-1}(\bar u) + \beta c_{G_{k-1}}(\bar u)$, and that $(1-\beta)c_{\text{inner}} \ge (1-\beta) c_{k-1}(u) + \beta c_{G_{k-1}}(u)$ for every other $u \in F(G_{k-1})$.

We can therefore obtain $p_k(v)$ from $p_{k-1}(v)$ as follows.
First, for $i=0,\ldots,k-2$ we rescale every $\mybeta{}{i}$ multiplying it by $(1-\beta)$.
Second, we let $\mybeta{}{k-1}=\beta$.
Third, we set $u_k = \bar u$.
Fourth, we set $\mybeta{}{k}=0$.
The expression of $p_k(v)$ still equals the expression of Equation~\ref{eqn:pkv_next} and, by the argument above, satisfies Equations~\ref{eqn:balanced_coefficients_inner} and \ref{eqn:balanced_coefficients_outer}.

Finally, note that the information necessary to compute $p_{k}(v)$ from $p_{k-1}(v)$ is obtained by expanding $u_k$ if we have already expanded $u_0,\ldots,u_{k-1}$.
Since expanding $u_i$ requires $1+2\,\indegree{u_i}$ queries, the total query complexity of building $p_k(v)$ is $\sum_{i=0}^k (1+ 2\,\indegree{u_i})$.

\subsection{Proof of Lemma~\ref{lem:vplus}}
\label{apx:vplus}
We shall analyse the rescaled random variable $S_u = \ell \cdot s(u)$ counting the number of samples yielding $u$.
If $P(u) \ge \frac{25\ln(2n/\delta)}{\epsilon^2 \ell}$, then $\E[S_u] \ge \frac{25\ln(2n/\delta)}{\epsilon^2}$ and the event $u \notin B$ corresponds to $S_u < \frac{16\ln(2n/\delta)}{ \epsilon^2} = (1-9/25)\E[S_u]$.
By the concentration bounds of Appendix~\ref{apx:chernoff_bounds}, $\prob[S_u < (1-9/25)\E[S_u]\big] < \exp\!\big({- \frac{(9/25)^2}{2} \frac{25\ln(2n/\delta)}{\epsilon^2}}\big) = (\frac{\delta}{2n})^{1.62} < \frac{\delta}{3n}$.

Consider now any $u \in B$.
By construction $S_u \ge \frac{16\ln(2n/\delta)}{ \epsilon^2}$.
If $S_u < (1-\epsilon)\E[S_u]$ then $\E[S_u] > \frac{16\ln(2n/\delta)}{ \epsilon^2(1-\epsilon)}$ and thus $\prob[S_u < (1-\epsilon)\E[S_u]] < \exp\!\big(-\frac{\epsilon^2}{2} \frac{16\ln(2n/\delta)}{ \epsilon^2(1-\epsilon)} \big) < (\frac{\delta}{2n})^{8} < \frac{\delta}{256n}$.
If instead $S_u > (1+\epsilon)\E[S_u]$ then $S_u = (1+\bar{\epsilon})\E[S_u]$ for some $\bar{\epsilon} \ge \epsilon$, and $\E[S_u] \ge \frac{16\ln(2n/\delta)}{\epsilon^2(1+\bar{\epsilon})}$.
Again by the bounds of Appendix~\ref{apx:chernoff_bounds} we get $\prob[S_u = (1+\bar{\epsilon})\E[S_u]] \le \exp\!\big(\!-\frac{\bar{\epsilon}^2}{3} \frac{16\ln(2n/\delta)}{\epsilon^2(1+\bar{\epsilon})} \big)$.
Now if $\bar{\epsilon} \ge 1$ then $\frac{\bar{\epsilon}^2}{1+\bar{\epsilon}} \ge \frac{1}{2}$, while if $\bar{\epsilon} < 1$ then $\frac{\bar{\epsilon}^2}{1+\bar{\epsilon}} \ge \frac{1}{2}$; and since $\epsilon \le 1$ and $\bar{\epsilon} \ge \epsilon$, in both cases $\frac{\bar{\epsilon}^2}{\epsilon^2(1+\bar{\epsilon})} \ge \frac{1}{2}$ and $\exp\!\big(\!-\frac{\bar{\epsilon}^2}{3} \frac{16\ln(2n/\delta)}{\epsilon^2(1+\bar{\epsilon})} \big) \le (\frac{\delta}{2n})^{16/6} < \frac{\delta}{6.34n}$.

By a union bound on all $u$, the probability that at least one of (i) and (ii) fails is at most $\frac{\delta}{3n} + \frac{\delta}{256n}+\frac{\delta}{6.34n} < \frac{\delta}{2}$.

\subsection{Proof of Lemma~\ref{lem:pc}}
\label{apx:indegineq}
Recall that $P(u) = \frac{1-\alpha}{n} + \alpha\sum_{w \rightarrow u} \frac{P(w)}{\outdegree{w}} \ge \frac{\alpha(1-\alpha)}{n}\sum_{w \rightarrow u} \frac{1}{\outdegree{w}}$.
On the one hand, since $\outdegree{w} \le \dmax$ this implies $P(u) = \Omega(\frac{\indegree{u}}{n \dmax})$ and thus $\indegree{u} = O(n \dmax P(u)) = O(\frac{m \dmax P(u)}{d})$.
%Since by hypothesis point (i) of Lemma~\ref{lem:vplus} holds, then for any $u_i$ we have $P(u_i) = O\big(\frac{\ln(n/\delta)}{\ell \epsilon^2}\big)$ and thus $\indegree{u_i} = O\big(\frac{\dmax n \ln(n/\delta)}{\ell \epsilon^2 }\big)$.
%It follows that $t = O\big(\frac{\dmax m n \ln(n/\delta)}{\ell \epsilon^2}\big)$.
On the other hand, by the harmonic--arithmetic mean inequality $\frac{1}{\indegree{u}}\sum_{w \rightarrow u} \outdegree{w} \ge \frac{\indegree{u}}{\sum_{w \rightarrow u}1/\outdegree{w}}$, that is, $\sum_{w \rightarrow u} \frac{1}{\outdegree{w}} \ge \frac{\indegree{u}^2}{\sum_{w \rightarrow u} \outdegree{w}}$.
This in turn implies $P(u) = \Omega\big(\frac{1}{n} \frac{\indegree{u}^2}{\sum_{w \rightarrow u} \outdegree{w}}\big)$.
But  $\sum_{w \rightarrow u} \outdegree{w} \le m$ and thus $P(u) = \Omega\big(\frac{1}{n} \frac{\indegree{u}^2}{m}\big)$, that is, $\indegree{u}  = O((m n P(u))^{1/2}) = O(\frac{m P(u)^{1/2}}{d^{1/2}})$.
%But point (i) of Lemma~\ref{lem:vplus} implies $P(u_i) = O\big(\frac{\ln(n/\delta)}{\ell \epsilon^2}\big)$ and thus $\indegree{u_i}  = O\big(\big(E n \frac{\ln(n/\delta)}{\ell \epsilon^2}\big)^{1/2}\big)$ for all $u_i$.
%Therefore $t = O\big(\frac{m}{\epsilon}\big( \frac{E n \ln(n/\delta)}{\ell} \big)^{\frac{1}{2}}\big)$.

\subsection{Proof of Lemma~\ref{lem:pmv_approx}}
\label{apx:pmv_approx}
We build a $p_k'(v)$ whose generic coefficient differs additively by at most $\frac{\bareps}{2}$ from its counterpart in $p_k(v)$.
It is easy to see that $p_k'(v)$ is an additive $\bareps$-approximation of $p_k(v)$.
Indeed, write $p_k(v)$ in the form $p_k(v) = c_k + \sum_{u \in G} c_k(u) \chi_u$, where $c_k(u)=0$ if $u$ does not appear in the expression of $p_k(v)$ given by Equation~\ref{eqn:p_m_new}.
Do the same for $p_k'(v)$.
Now suppose $|c_k-c_k'| \le \frac{\bareps}{2}$ and that $|c_k(u)-c_k'(u)| \le \frac{\bareps}{2}$ for all $u \in G$.
Then $|\E[p_k(v) - p_k'(v)]| \le |c_k-c_k'|+ \sum_{u \in G} |c_k(u)-c_k'(u)| \E[\chi_u] \le \frac{\bareps}{2} + \sum_{u \in G}\frac{\bareps}{2}\E[\chi_u] = \bareps$, where we used the fact that $\sum_{u \in G}\E[\chi_u] = 1$.
%For (1), consider the coefficient $c_k'$.
%If $c_k' < \bareps$, then since $c_k' \ge c_k - \bareps^2$ we have $c_k \le \bareps + \bareps^2 \le 2\bareps$.
%If instead $c_k' \ge \bareps$, then $c_k \ge \bareps$ too, and since $c_k' \ge c_k - \bareps^2$ we have $c_k' \ge c_k(1 - \bareps)$.
%This argument applies to every other coefficient as well.
%Therefore the difference $\E[p_k'(v)] - \E[p_k(v)]$ restricted to the terms that in $p_k'(v)$ have a coefficient $< \bareps$ is bounded from below by $-4\epsilon$:
%%Consider now the expectation of $p_k'(v)$ restricted to the terms with coefficient smaller than $\bareps$.
%%Such an expectation is at most $3\bareps$:
%at most $-2\bareps$ comes from $c_k'-c_k$ (in the case $c_k' \le \bareps$), and at most $\E[\sum_{u \in G} \chi_u \cdot 2\bareps] = 2\bareps$ comes from other coefficients.
%On the other hand, $\E[p_k'(v)]$ restricted to the terms that in $p_k'(v)$ have a coefficient $\ge\bareps$ is at least $(1-\bareps)$ times its counterpart in $\E[p_k(v)]$.
%Therefore $\E[p_k'(v)] \ge \E[p_k(v)](1-\bareps) - 4\bareps$.
%The other side of the inequality is trivial.

Let us then prove the claim for $p_k'(v)$, by induction.
At each step we proceed as with $p_k(v)$ (see Appendix~\ref{apx:construction}), but, instead of the coefficients $c_{G_k}(u)$, we use additive $\frac{\bareps}{2}$-approximations $c'_{G_k}(u)$ i.e.\ such that $|c_{G_k}(u)-c'_{G_k}(u)| \le \frac{\bareps}{2}$.
This suffices, since $c_k'$, $c_k'(u_i)$, and $c_k'(u)$ are weighted averages of the $c_{G_k}'(u)$ (see Section~\ref{sub:exploring_choosing}) and thus are $\frac{\bareps}{2}$-approximations of $c_k$, $c_k(u_i)$, and $c_k(u)$.
Assume we have built $p_{k-1}'(v)$ for some $k-1 \ge 1$.
(The analysis holds trivially for $k-1 \le 1$, too).
%To build $p_k'(v)$ from $p_{k-1}'(v)$ we proceed as usual .
%Furthermore, once we have $c_{G_k}$, $c_{G_k}(u_i)$, and $c_{G_k}(u)$, we can compute the coefficients $c_k$, $c_k(u_i)$ and $c_k(u)$ in $O(t)$ operations.
To build $p_k'(v)$ we must: (A) choose $u_k$ and $\mybeta{}{k-1}$, (B) compute $c_{G_k}'$ and $c_{G_k}'(u)$ for all $u \in F(G_k)$, and (C) compute the new coefficients $c_k'$, $c_k'(u_i)$, and $c_k'(u)$.
Note that $t$ is an upper bound to the number of arcs and/or nodes in $G_{k-1}$ and $F(G_{k-1})$.
For (A), we use Equation~\ref{eqn:beta_oneshot} from Appendix~\ref{apx:construction}, which requires $O(|F(G_{k-1})|)=O(t)$ operations.
For (C), once we know the $c_{G_k}'(u)$, we use Equation~\ref{eqn:pkv_next}, which again requires $O(|F(G_{k-1})|+|G_{k-1}|) = O(t)$ operations.

Let us then address (B), showing how to compute $c_{G_k}'$ and $c_{G_k}'(u)$ for all $u \in F(G)$ in time $O(t \ln(1/\bareps))$.
First note that $\coefc{G_k}$ and $\coefu{G_k}{u}$ are bounded by weighted averages of resistances $\cond{G_k}{u_i}{v}$ for $u_i \in G_k$ (see Appendix~\ref{apx:PR_conductance} and~\ref{apx:definitions}).
Therefore we just need an additive $\frac{\bareps}{2}$-approximation of each $\cond{G_k}{u_i}{v}$.
%Let us analyse the steps taken by BalancedEstimator.
%Recall from Appendix~\ref{apx:PR_conductance} and~\ref{apx:definitions} that $c_{G_k} = \sum_{w:(u,w)\in F(G_k)} \frac{\alpha}{\outdegree{u}}\,\cond{G_k}{w}{v}$, where:
%%
%% $c_{G_k}(u)=$ is the sum of the resistances of all paths from $u$ to $v$ through $G_k$.
%%Since the resistance of paths of length $\tau$ is bounded by $\alpha^\tau$, we can just disregard paths longer than $a \ln(1/\bareps)$ for a large enough $a > 0$.
%%From Appendix~\ref{apx:PR_conductance}:
%\begin{align}
%\cond{G_k}{u}{v} &= \sum_{\tau \geq 0} \alpha^{\tau}  \sum_{\substack{\pi \in \Pi_{G_k}(u,v) \\ |\pi|=\tau}} \,\, \prod_{(w,w')\in \pi}\frac{1}{\outdegree{w}}
%\end{align}
%where $\Pi_{G_k}(u,v)$ is the set of paths in $G_k$ from $u$ to $v$.
%We can express the relatonships between the $\cond{G_m}{u}{v}$ more compactly in matricial form.
%Let us restate the equation more succintly in matricial form.
For any $\tau \ge 0$ let:
\begin{align}
r_k^\tau(u_i) = \alpha^{\tau}  \sum_{\substack{\pi \in \Pi_{G_k}(u_i,v) \\ |\pi|=\tau}} \,\, \prod_{(w,w')\in \pi}\frac{1}{\outdegree{w}}
\end{align}
One can see that $\cond{G_k}{u_i}{v} = \sum_{\tau \geq 0} r_k^\tau(u_i)$ (see Appendix~\ref{apx:definitions}).
%; this follows by seeing each path as the arc $(w,u)$ followed by a path of $k-1$ arcs from $u$ to $v$.
%Recall that the nodes of $G_m$ are $u_0,\ldots,u_m$.
%Let now $\mathbf{r}_m^i \in \mathbb{R}^m$ be the vector whose $i$-th component is $\cond{G_m}{u_i}{v}$.
%Note that 
For all $\tau \ge 0$ let $\mathbf{r}_k^\tau$ be the vector whose $i$-th component is $r_k^\tau(u_i)$, for $i=0,\ldots,k$ (for $\tau=0$ we set $r_k^0(v) = 1$ and $r_k^0(u_i)=0$ for $u_i \ne v$).
Let $\bfA_k \in \mathbb{R}^{(k+1)\times(k+1)}$ be the normalized adjacency matrix of $G$ restricted to $G_k$, so $\bfA_k[i,j] = \frac{1}{\outdegree{u_i}}$ if and only if $u_i,u_j \in G_k$ and $(u_i,u_j) \in G$.
Then for all $\tau \ge 1$:
\begin{align}
\label{eq:paths_matrix_series}
\mathbf{r}_k^\tau = \alpha\, \bfA_k \, \mathbf{r}_k^{\tau-1}
\end{align}
We can thus compute all $\cond{G_k}{u_i}{v}$ simultaneously as $\mathbf{r}_k =  \sum_{\tau \geq 0} \alpha^t (\bfA_k)^\tau \mathbf{r}_k^{0}$.
But $\bfA_k$ has at most $t$ non-zero entries, so computing $\alpha \bfA_k \mathbf{r}_k^{\tau-1}$ from $\bfA_k$ and $\mathbf{r}_k^{\tau-1}$ by sparse matrix-vector product takes time $O(t)$.
Since $r_k^\tau(u_i) \le \alpha^\tau$, if we pick $\bar{\tau} = a \ln(1 / \bareps)$ for a large enough $a > 0$ and we let $\mathbf{r}_k' = \sum_{\tau = 0}^{\bar{\tau}} \mathbf{r}_k^\tau$, it holds:
\begin{align}
\label{eq:paths_matrix_series_tail}
\mathbf{r}_k' = \mathbf{r}_k - \sum_{\tau > \bar{\tau}+1} \mathbf{r}_k^\tau \ge \mathbf{r}_k - \frac{\alpha^{\bar{\tau}+1}}{1-\alpha}\cdot \mathbf{1} \ge \mathbf{r}_k - \frac{\bareps}{2}
\end{align}
Thus $\mathbf{r}_k'$ gives additive $\frac{\bareps}{2}$-approximations of $\cond{G_k}{u_i}{v}$, and can be computed in time $O(t \ln(1 / \bareps)$.
%Finally, we show that using $\mathbf{r}_k'$ we get an additive $\frac{\bareps}{2}$-approximations of $\coefc{G_k}$ and $\coefu{G_k}{u}$.
%By definition, $\coefc{G_k} = \sum_{w \in G_k} \!\frac{1-\alpha}{n}\, \cond{G_k}{w}{v}$ and $\coefu{G_k}{u} = \sum_{w:(u,w)\in F(G_k)} \, \frac{\alpha}{\outdegree{u}} \cond{G_k}{w}{v}$ (see Appendix~\ref{apx:PR_conductance}).
%By using the $\mathbf{r}_m'$ we then get additive $\bareps^2$-approximations of $\coefc{G_m}$ and $\coefu{G_m}{u}$.
%In conclusion, the complexity of building $p_{k}'(v)$ from $p_{k-1}'(v)$ is $O(t \ln(1 / \bareps)$.
Summing over $k$ steps concludes the proof.

\subsection{Dangling nodes}
\label{apx:dangling}
\newcommand{\ev}{\mathcal{E}}
Essentially, if $G$ contains dangling nodes then the estimators for $P(v)$ developed in Section~\ref{sec:upperbounds} must be adapted by adding a random term (and, possibly, rescaling by a constant) which can be estimated with sufficient accuracy through just $O(1)$ calls to \samplenode{}.
In what follows we let $G_\emptyset = \{u \in G : \outdegree{u}=0\}$ and $P_{\emptyset} = \sum_{u\in G_\emptyset \setminus v} P(u)$.

%We start by generalizing Equation~\ref{eqn:PRmho} to allow for dangling nodes in $G$ -- but not in $\bar{G} \setminus \{v\}$.
%This is the most general assumption we need, since the subgraph $\bar{G}$ built by our algorithm never contains dangling nodes except possibly $v$.
\begin{lemma}
\label{lem:PR_conductance_general}
For any induced subgraph $H$ of $G$ and any $v \in H$, if $H \setminus v$ is free from dangling nodes then $P(v)$ has the form:
\begin{align}
P(v) = \mu_{H} \Big(c_H \big(1+\frac{\alpha}{1-\alpha} P_{\emptyset}\big) + \sum_{u \in F(H)} \!\!\! P(u) \cdot c_{H}(u)  \Big)
\end{align}
$\mu_{H}=(1 - \frac{\alpha}{1-\alpha}c_H)^{-1}$ if $v\in G_\emptyset$ and $\mu_{H}=1$ otherwise.
\end{lemma}
\begin{proof}
%We shall show that:
%\begin{align}
%P(v) = \mu_{H}
%  \Big( \sum_{z\in H}
%       \frac{1-\alpha}{n} \cond{H}{z}{v}
%       + \sum_{z\in H}\sum_{u\in G_\emptyset \setminus v}
%       \!\!\!\! P(u) \frac{\alpha}{n}  \cond{H}{z}{v}
%  +\!\!\!\!\sum_{(u,w)\in F(H)}\!\!\!\!\!\! P(u) \frac{\alpha}{\outdegree{u}}
%       \cond{H}{w}{v}
%  \Big)
%\end{align}
%where $\mu_{H}=(1 - \frac{\alpha}{n}\sum_{z\in H} \cond{H}{z}{v})^{-1}$ if $v$ is dangling and $\mu_{H}=1$ otherwise.
%The claim follows by setting $c_H$ and $c_H(u)$ as in Appendix~\ref{apx:PR_conductance}.
%We extend the proof of Equation~\ref{eqn:PRmho} (see Appendix~\ref{apx:PR_conductance}).
%For the sake of the analysis we imagine that, if the current node is dangling, with probability $1-\alpha$ the walk performs the usual random jump and with probability $\alpha$ it follows one at random of $n$ outgoing virtual arcs.
We shall adapt Equation~\ref{eqn:Pv_mho} from Appendix~\ref{apx:PR_conductance}.

Consider first the case $v \notin G_{\emptyset}$.
In this case we must add to the second term of the right-hand side of Equation~\ref{eqn:Pv_mho} the paths containing nodes of $G_\emptyset = G_{\emptyset} \setminus v$.
We can as usual break each path into a head path $\pi'$ terminating in a node $u \in F(H)$ and a tail path $\pi' \in H$, joined by an arc of $F(H)$; and then gathering terms according to $u \in F(H)$.
However, since each $u \in G_\emptyset$ is virtually on the frontier $F(H)$, this amounts just to adding terms $P(u)\frac{\alpha}{\outdegree{u}}\cond{H}{z}{v}$ for all $u \in G_\emptyset$ and all $z \in H$ -- all paths having $u \notin G_\emptyset$ are already in the expression of Equation~\ref{eqn:Pv_mho}.
Since $\outdegree{u}=n$ and $G_\emptyset = G_\emptyset \setminus v$, the whole expression becomes $\sum_{z\in H} \sum_{u\in G_\emptyset \setminus v} P(u) \frac{\alpha}{n} \cond{H}{z}{v}$, which is equivalent to $P_{\emptyset} \frac{\alpha}{n} \sum_{z\in H} \cond{H}{z}{v} = P_{\emptyset} \frac{\alpha}{1-\alpha}c_H$.
%We get:
%\begin{align}
%\label{eqn:dangling1}
%P(v) = 
%  \Big( \sum_{z\in H}
%       \frac{1-\alpha}{n} \cond{H}{z}{v}
%       + \sum_{z\in H}\sum_{u\in G_\emptyset \setminus v}
%       \!\!\!\! P(u) \frac{\alpha}{n}  \cond{H}{z}{v}
%  +\!\!\!\!\sum_{(u,w)\in F(H)}\!\!\!\!\!\! P(u) \frac{\alpha}{\outdegree{u}}
%       \cond{H}{w}{v}
%  \Big)
%\end{align}
We obtain:
\begin{align}
\label{eqn:dangling1}
P(v) = c_H  + \sum_{u \in F(H)} \!\!\! P(u) \cdot c_{H}(u) + \frac{\alpha}{1-\alpha} P_{\emptyset}
\end{align}
Reordering terms and multiplying by $\mu_{H}=1$ concludes this case.

Now suppose instead $v \in G_{\emptyset}$.
In this case we must add to the previous case the walks containing $v$ as intermediate node.
%Note that the preious case already takes into account all walks \emph{not} going through $v$
More precisely, we add only those paths of the form $(\pi',\pi,v)$ where $\pi'$ ends in $v$ and such that $\pi \in H \setminus v$; since all paths of this form but where $\pi$ contains some $u \in G_\emptyset \setminus v$ are already counted by the case $v \notin G_{\emptyset}$.
As usual, we observe that summing over all $\pi'$ just gives $P(v)$, and summing over all $(\pi,v)$ such that $\pi \in H \setminus v$ gives $\sum_{z\in H} \cond{H}{z}{v}$.
We shall then multiply by the usual factor $\frac{\alpha}{n}$ for taking one of the outgoing arcs of $v$.
We are therefore adding to the right-hand side of Equation~\ref{eqn:dangling1} the term:
\begin{align}
\label{eqn:inpath_dangling}
P(v) \frac{\alpha}{n} \sum_{z\in H} \cond{H}{z}{v} = P(v) \frac{\alpha}{1-\alpha}c_H
\end{align}
Thus $P(v)$ equals the right-hand side of Equation~\ref{eqn:dangling1} multiplied by $(1-\frac{\alpha}{1-\alpha}c_H)^{-1}$.
\end{proof}
Note that we can compute $(1-\frac{\alpha}{1-\alpha}c_H)^{-1}$ and thus $\mu_{H}$ deterministically.
Note also that we just need an accurate enough approximation of $(1 + \frac{\alpha}{1-\alpha} P_{\emptyset})$, e.g.\ within a multiplicative factor $1\pm O(\epsilon)$ with probability $1-O(\delta)$.
Let then $p_\emptyset = \frac{1}{\ell}\sum_{j=1}^{\ell}\sum_{u\in G_\emptyset \setminus v} \chi^j_u$.
Clearly $\E[p_\emptyset] = P_{\emptyset}$.
%, and we use $(1 + \frac{\alpha}{1-\alpha}p_\emptyset)$ as an estimator of $(1 + \frac{\alpha}{1-\alpha} P_{\emptyset})$.
Now, if $(1 + \frac{\alpha}{1-\alpha}p_\emptyset)$ falls off its expectation by more than a factor $1 \pm \epsilon$, then $\ell \, p_\emptyset$ falls off its expectation by more than a factor $1 \pm \epsilon\big(1 + \frac{1-\alpha}{\alpha P_\emptyset}\big)$.
Since $\ell \, p_\emptyset$ is a sum of non-positively correlated indicator random variables with expectation $\ell\, P_\emptyset$, by the probability bounds of Appendix~\ref{apx:chernoff_bounds} the probability that such an event takes place is at most
\begin{align}
2 \exp\Big(- \frac{\epsilon^2 (1 + \frac{1-\alpha}{\alpha P_\emptyset})^2}{3} \ell \, P_\emptyset\Big)
\le 
2 \exp\Big(- \frac{\epsilon^2(1-\alpha)^2\ell}{3\alpha^2} \Big)
\end{align}
Therefore it suffices to take $\ell = O\big(\frac{1}{\epsilon^2}\ln(\frac{2}{\delta})\big)$ additional samples.

\subsection{Heat kernel}
\label{apx:me}
In this section we adapt the algorithms, proofs, and bounds of Section~\ref{sec:upperbounds} to the case of heat kernel.
Recall (Equation~\ref{eqn:me_def}) that:
\begin{align}
\label{eqn:sc1}
P(v) = \frac{\eta}{n} \sum_{\tau \ge 0} \, \sum_{z \in G} \, \sum_{\substack{\pi \in \Pi_{G}(z,v) \\ |\pi| = \tau }} \!\!\!  \frac{\alpha^{\tau} \rho_{\pi}}{\tau!},
\qquad \eta = (\sum_{\tau\ge 0} \frac{\alpha^{\tau}}{\tau!})^{-1} = e^{-\alpha}
\end{align}
Unlike PageRank, we cannot express $P(v)$ directly as a function of $P(u), u \rightarrow v$ unless we break $P(u)$ over paths of different lengths.
Formally, we need to define:
\begin{align}
\label{eqn:ptau}
P_\tau(v) = \frac{\eta}{n} \sum_{z \in G} \,\sum_{\substack{\pi \in \Pi_{G}(z,v) \\ |\pi|=\tau}} \!\!\!  \frac{\alpha^{\tau} \rho_{\pi}}{\tau!}
\end{align}
Note that $P_0(u)=\frac{\eta}{n}$, while $P_\tau(u) = \sum_{w \rightarrow u} \frac{\alpha}{\tau \, \outdegree{w}} P_{\tau-1}(w)$ for $\tau \ge 1$; finally, $P(u) = \sum_{\tau \ge 0} P_\tau(u)$.
\\[5pt]\noindent \textbf{Remark.} Since $P(v) = \Omega(n^{-1})$ and $P_{\tau}(v) \le \frac{\alpha^\tau}{\tau!}$, we can disregard all $\tau \ge a \ln(n/\epsilon)$ for some sufficiently large $a \ge 1$ and still keep our approximation guarantees valid.
%From a computational point of view we therefore assume implicitly all such $\tau$ are disregarded.
Hence in what follows we implicitly consider $0 \le \tau < a\ln(n/\epsilon)$ even if, for readability, we write $\tau \ge 0$.
%The meaning will be clear from the context.

\subsubsection{Random walk sampling}
\label{sub:rwalk2}
As a first ingredient, we need to sample nodes $u \in G$ from the distributions given by $P_\tau(u)$.
For each $\tau \ge 0$ we do the following.
First, pick a node $z$ uniformly at random in $G$.
Then, walk $\tau$ steps from $z$, and let $\chi_{\tau,u}$ be the indicator random variable of the event that the random walk ends in $u$.
It is clear that $\E[\chi_{\tau,u}] = \frac{1}{n} \sum_{z \in G} \sum_{\pi \in \Pi_{G}(z,v) : |\pi|=\tau} \rho_{\pi} = \frac{\tau!}{\eta \alpha^\tau} P_\tau(u)$.
Note that taking one sample of $\chi_{\tau,u}$ requires $O(\ln(n/\epsilon))$ queries and elementary operations, since $\tau < a \ln(n/\epsilon)$.
%We then define $\chi_u = \eta \sum_{\tau \ge 0} \frac{\alpha^\tau}{\tau!} \chi_{\tau,u}$.
%Clearly $\chi_u \in [0,1]$ and $\E[\chi_u] = P(u)$.
%Taking one sample of $\chi_u$ has $O(\ln^2(n/\epsilon))$ query and computational complexity.
%Now let $\chi_{\tau,u}$ be the indicator random variable of the event that the random walk ends in $u$ and $L=\tau$.
%It is straightforward to see that $\E[\chi_{\tau,u}] = P_\tau(u)$.

\subsubsection{Subgraph estimators}
\label{sub:subgraph2}
%We shall adapt the subgraph estimator of PageRank to our case.
%The point is once again to break the score of $v$'s parents by path length.
We adapt the PageRank subgraph estimator by breaking the $P(u)$ by path length.
Formally:
\begin{lemma}
\label{lem:pr_diffuse_2}
For any induced subgraph $H$ of $G$ and any $v \in H$ it holds:
\begin{equation}
\label{eqn:PRmhox_2}
P(v) = \coefc{H} + \sum_{u \in F(H)} \sum_{\tau \ge 0} \, P_{\tau}(u) \frac{\tau!}{\eta \alpha^\tau} \cdot c_{H,\tau}(u)
\end{equation}
where $\coefc{H}$ and $c_{H,\tau}(u)$ depend only on $\tau$, $H$, $F(H)$ and on the outdegrees of $u \in H$ and $u \in F(H)$.
\end{lemma}
%\begin{lemma}
%For any induced subgraph $H$ of $G$ and any $v \in H$ it holds:
%\begin{align}
%\label{eqn:phv_2}
%p_H(v) = c_H + \sum_{u \in F(H)} \sum_{\tau \ge 0}  \,\chi_{\tau,u} \cdot c_{H,\tau}(u)
%\end{align}
%where $\coefc{H}$ and $c_{H,\tau}(u)$ depend only on $H$, $F(H)$ and on the outdegrees of $u \in H$ and $u \in F(H)$.
%\end{lemma}
\begin{proof}
As done in Section~\ref{apx:PR_conductance} for PageRank, we take Equation~\ref{eqn:sc1} and consider separately paths in $H$ and paths containing an arc of $F(H)$, breaking these latter on that arc.
This gives:
\begin{align}
\label{eqn:sc2}
P(v) = \frac{\eta}{n} \sum_{z \in H} \sum_{\pi \in \Pi_{H}(z,v)} \!\!\!\! \frac{\alpha^{|\pi|} \rho_{\pi}}{|\pi|!}
+ 
\frac{\eta}{n} \sum_{(u,w) \in F(H)} \sum_{z \in G} \, \sum_{\pi \in \Pi_{G}(z,u)} \sum_{\pi' \in \Pi_{H}(w,v)}  \frac{\alpha^{|\pi| + |\pi'| + 1} \rho_\pi \rho_{\pi'} }{\outdegree{u} (|\pi| + |\pi'| + 1)!}
\end{align}
We let $c_H = \frac{\eta}{n}\sum_{z \in H} \sum_{\pi \in \Pi_{H}(z,v)} \frac{\alpha^{|\pi|} \rho_{\pi}}{|\pi|!}$.
For the second term, we group the outmost summation by $u$ and break $\Pi_{G}(z,u)$ over all possible lengths $\tau \ge 0$.
After simple rearrangements we get:
\begin{align}
\label{eqn:sc3}
\sum_{u \in F(H)} \sum_{\tau \ge 0} \Big( \frac{1}{n} \sum_{z \in G} \sum_{\substack{\pi \in \Pi_{G}(z,u) \\ |\pi|=\tau}} \rho_{\pi} \Big)
\Big( \eta \sum_{w:(u,w) \in F(H)} \sum_{\pi' \in \Pi_{H}(w,v)}  \frac{\rho_{\pi'} \, \alpha^{\tau + |\pi'| + 1}}{\outdegree{u} (\tau + |\pi'| + 1)!} \Big) 
\end{align}
The first factor inside the summation is exactly $P_\tau(u) \, \frac{\tau!}{\eta \alpha^\tau}$ (see Equation~\ref{eqn:ptau}).
%\begin{align}
%\label{eqn:sc3}
%\sum_{u \in F(H)} \sum_{\tau \ge 0}
%P_\tau(u) \, \frac{\tau!}{\eta \alpha^\tau}
%\cdot
%\Big( \eta \sum_{w:(u,w) \in F(H)} \sum_{\pi' \in \Pi_{H}(w,v)}  \frac{\rho_{\pi'} \, \alpha^{\tau + |\pi'| + 1}}{\outdegree{u} (\tau + |\pi'| + 1)!} \Big) 
%\end{align}
Now define: %let $c_{H,\tau}(u)$ be the term between brackets,
\begin{align}
\label{eqn:c_h_tau}
c_{H,\tau}(u) = \eta \sum_{w:(u,w) \in F(H)} \sum_{\pi' \in \Pi_{H}(w,v)}  \frac{\rho_{\pi'} \, \alpha^{\tau + |\pi'| + 1}}{\outdegree{u} (\tau + |\pi'| + 1)!}
\end{align}
and by Equations~\ref{eqn:sc2}-\ref{eqn:sc3} the claim is proven.
\end{proof}
\noindent By replacing $P_\tau(u) \, \frac{\tau!}{\eta \alpha^\tau} = \E[\chi_{\tau,u}]$ with $\chi_{\tau,u}$, we obtain:
\begin{definition}
 \label{def:diffuse_2}
The \emph{subgraph estimator} of $P(v)$ given by $H$ is the random variable:
\begin{align}
\label{eqn:phv_2}
p_{H}(v) = \coefc{H} + \sum_{u \in F(H)} \sum_{\tau \ge 0}  \,\chi_{\tau,u} \cdot c_{H,\tau}(u)
\end{align}
\end{definition}

%\noindent \textbf{Note.} For $\tau=0$ we can actually replace $\chi_{\tau,u}$ with $\frac{1}{n}$, since we know $\E[\chi_{0,u}]=\frac{1}{n}$.
%We will do this later on.

\subsubsection{Weighted estimators}
The definitions and arguments of Section~\ref{sub:weighted} can be adapted straightforwardly.
The only change is in the final form of the estimator, which is given by a variation of Equation~\ref{eqn:p_m_2}:
\begin{align}
\label{eqn:p_m_2_2}
p_k(v) = \sum_{i = 0}^{k} \mybeta{}{i} \, \coefc{G_i} 
+ \sum_{i = 1}^{k} \, \sum_{\tau \ge 0} \chi_{\tau,u_i} \cdot \!\!\sum_{j=j(u_i)}^{i-1} \!\! \mybeta{}{j}\, \coefu{G_j,\tau}{u_i}
 + \!\!\!\sum_{u \in F(G_{k})} \sum_{\tau \ge 0} \chi_{\tau,u} \cdot \!\!\sum_{j = j(u)}^{k} \!\! \mybeta{}{j}\, \coefu{G_j,\tau}{u}
\end{align}

\subsubsection{Building a perfect weighted estimator}
Similarly to PageRank, define $c_k = \sum_{i = 0}^{k} \mybeta{}{i} \, \coefc{G_i}$, define $c_{k,\tau}(u_i) = \sum_{j=j(u_i)}^{i-1} \mybeta{}{j}\, \coefu{G_j,\tau}{u_i}$, and define $c_{k,\tau}(u) = \sum_{j = j(u)}^{k} \mybeta{}{j}\, \coefu{G_j,\tau}{u}$.
Equation~\ref{eqn:p_m_2_2} becomes:
%If moreover we bring terms associated to the $\chi_{0,u_i}$ on their own, from Equation~\ref{eqn:p_m_2_2} we get:
\begin{align}
\label{eqn:p_m_new_2_apx}
%p_k(v) = c_k + \sum_{i = 1}^{k} \chi_{0,u_i} \, c_{k,0}(u_i) \,+\, \sum_{i = 1}^{k} \sum_{\tau \ge 1} \chi_{\tau,u_i} \, c_{k,\tau}(u_i) \,+\, \!\sum_{u \in F(G_{k})} \sum_{\tau \ge 0} \chi_{\tau,u} \, c_{k,\tau}(u)
p_k(v) = c_k + \sum_{i = 1}^{k} \sum_{\tau \ge 0} \chi_{\tau,u_i} \, c_{k,\tau}(u_i) \,+\, \!\sum_{u \in F(G_{k})} \sum_{\tau \ge 0} \chi_{\tau,u} \, c_{k,\tau}(u)
\end{align}
Now a crucial observation.
From Equation~\ref{eqn:c_h_tau}, and since $\frac{\alpha^{|\pi'| + 1}}{(|\pi'| + 1)!} \ge \frac{\alpha^{\tau + |\pi'| + 1}}{(\tau + |\pi'| + 1)!}$, for any given $H,u$ we have $c_{H,0}(u) \ge c_{H,\tau}(u)$ for all $\tau \ge 0$.
This implies $c_{k,0}(u_i) \ge c_{k,\tau}(u_i)$ for all $i=1,\ldots,k$ and $c_{k,0}(u) \ge c_{k,\tau}(u)$ for all $u \in F(G_k)$.
In other words the coefficients for $\tau=0$ dominate.
We then build $p_k(v)$ as done for PageRank (Section~\ref{sub:exploring_choosing} and Appendix~\ref{apx:construction}), but looking only at the coefficients $c_{k,0}(u_i)$ and $c_{k,0}(u)$.
We get a perfect weighted estimator in the following sense:
\begin{definition}
\label{def:balanced_estimator_2}
We say the weighted estimator $p_k(v)$ is \emph{perfect} if:
\begin{align*}
\begin{array}{ll}
c_{k,0}(u_i) = c_{k,0}(u_{i'}) &\text{for all } i,i' \in \{1,\ldots,k\}\\
%c_{k,0}(u_i) \ge c_{k,\tau}(u_i) &\text{for all } i \in \{1,\ldots,k\}\\
c_{k,0}(u_i) \ge c_{k,0}(u) &\text{for all } i \in \{1,\ldots,k\} \text{ and all } u \in F(G_k)
\end{array}
\end{align*}
\end{definition}
\noindent By straightforwardly adapting the proof of Lemma~\ref{lem:building}, one proves:
\begin{lemma}
\label{lem:building2}
We can build a perfect weighted estimator $p_k(v)$ using $\sum_{i=0}^k (1+ 2\,\indegree{u_i})$ queries.
\end{lemma}

\subsubsection{Concentration bounds}
\label{sub:conc2}
Let $p_k(v)$ be a perfect weighted estimator according to Definition~\ref{def:balanced_estimator_2}.
Recall (see Equation~\ref{eqn:p_m_new_2_apx}) that $c_{k,0}(u_i) \ge c_{k,\tau}(u_i)$ and $c_{k,0}(u) \ge c_{k,\tau}(u)$.
Then let $c=c_{k,0}(u_i)$ and consider the random variable $c^{-1}(p_k(v) - c_k)$, which is a sum of non-positively correlated binary random variables with coefficients in $[0,1]$.
The arguments of Section~\ref{sub:conc} apply, and Equation~\ref{eqn:pml_bound_2} holds.
Moreover $c^{-1}(p_k(v) - c_k) \ge \sum_{i=1}^k \chi_{0,u_i}$ and $\E[\sum_{i=1}^k \chi_{0,u_i}] = k\frac{1}{n}$, therefore $\E[c^{-1}(p_k(v) - c_k)] \ge \frac{k}{n}$.
Similarly to PageRank, by averaging over $\ell$ independent samples $\chi_{\tau,u}^1,\ldots,\chi_{\tau,u}^\ell$ of the $\chi_{\tau,u}$ for each $\tau$ we get an estimator $p_k^\ell(v)$ such that:
\begin{align}
\label{eqn:conc_pkl_2}
\prob\big[\,|p_k^{\ell}(v) - P(v)| > \epsilon P(v)\big] \leq 2\exp\!\Big(\!-\frac{\epsilon^2 k \ell }{3n}\Big)
\end{align}
We thus need $\ell = O(\frac{n}{k} \epsilon^{-2} \ln(1/\delta))$, as for PageRank.
Note however that, unlike PageRank, the query and computational complexity of sampling $p_k^{\ell}(v)$ is $\Theta(\ell \ln^2(n/\epsilon))$, since for each $\tau = 0,\ldots, a \ln(n/\epsilon)-1$ we take $\ell$ samples and each sample costs $O(\ln(n/\epsilon))$.

\subsubsection{Blacklisting heavy nodes}
We adapt Section~\ref{sub:blacklisting} as follows.
For each $\tau \ge 0$ take again $\ell$ independent samples $\chi_{\tau,u}^1,\ldots,\chi_{\tau,u}^\ell$ of the $\chi_{u,\tau}$.
For any given $H$ and any $u \in F(H)$ define:
\begin{align}
\label{eqn:shu}
s_H(u) = \sum_{\tau \ge 1} \Big(\frac{1}{\ell} \sum_{j=1}^\ell \chi_{\tau,u}^j\Big) \cdot \frac{c_{H,\tau}(u)}{c_{H,1}(u)}, \qquad \sigma_H(u) = \E[s_H(u)]
\end{align}
Now let $B = \{u \in G : \frac{1}{\ell} \sum_{j=1}^\ell \chi_{1,u}^j \ge \frac{16\ln(2n/\delta)}{\epsilon^2\ell}\}$.
We prove:
\begin{lemma}
\label{lem:vplus2}
With probability at least $1-\frac{\delta}{2}$ it holds:\\
\hspace*{10pt} \emph{1.} $\{u \in G : P_1(u) \ge \frac{25 \eta \alpha \ln(2n /\delta)}{\epsilon^2 \ell}\} \subseteq B$\\[2pt]
\hspace*{10pt} \emph{2.} $(1 - \epsilon )\sigma_H(u) \le  s_H(u) \le (1 - \epsilon )\sigma_H(u)$ for all $u \in B$ and all $H \subseteq G$
\end{lemma}
\begin{proof}
We adapt the proof of Lemma~\ref{lem:vplus} (Appendix~\ref{apx:vplus}).
For (1), if $P_1(u) \ge \frac{25 \eta \alpha \ln(2n /\delta)}{\epsilon^2 \ell}$ then $\E[\chi_{1,u}] \ge \frac{25 \ln(2n /\delta)}{\epsilon^2 \ell}$ and thus $\E[\sum_{j=1}^\ell \chi_{1,u}^j] \ge \frac{25 \ln(2n /\delta)}{\epsilon^2}$.
%But hence $\E[S_u] \ge \frac{25 \ln(2n /\delta)}{\epsilon^2}$ and 
The event $u \notin B$ instead implies that $\sum_{j=1}^\ell \chi_{1,u}^j < \frac{16\ln(2n/\delta)}{\epsilon^2}$.
As in the proof of Lemma~\ref{lem:vplus} we then get $\prob[u \notin B] < \frac{\delta}{3n}$.
%For (1), $\ge \frac{25\ln(2n /\delta)}{\epsilon^2 \ell}$
For (2), we shall analyse the rescaled random variable $S_u = \ell \cdot s_H(u) = \sum_{\tau \ge 1} \big(\sum_{j=1}^\ell \chi_{\tau,u}^j\big) \cdot \frac{c_{H,\tau}(u)}{c_{H,1}(u)}$, which is a sum of non-positively correlated binary random variables with coefficients in $[0,1]$ since $c_{H,\tau+1} < c_{H,\tau}$ (see Equation~\ref{eqn:c_h_tau}).
By construction $S_u \ge \sum_{j=1}^\ell \chi_{1,u}^j$, which means for all $u \in B$ we have $S_u \ge \frac{16\ln(2n/\delta)}{\epsilon^2}$.
Note this holds \emph{independently} of $H$.
We can now apply the argument of the proof of Lemma~\ref{lem:vplus} to $S_u$.
By a union bound on all $u \in G$ we get the thesis.
\end{proof}

We then proceed as for PageRank, ignoring the nodes in $B$ on the frontier during the construction of $p_k(v)$.
By Equation~\ref{eqn:p_m_new_2_apx}, the resulting estimator $q_k(v)$ has the form:
\begin{align}
\label{eqn:p_m_new_3}
q_k(v) = c_k + \sum_{i = 1}^{k} \sum_{\tau \ge 0} \chi_{\tau,u_i} \, c_{k,\tau}(u_i) \,+\, \!\sum_{\substack{u \in F(G_{k}) \\ u \notin B}} \sum_{\tau \ge 0} \chi_{\tau,u} \, c_{k,\tau}(u) \,+\, \!\sum_{\substack{u \in F(G_{k}) \\ u \in B}} \sum_{\tau \ge 0} \chi_{\tau,u} \, c_{k,\tau}(u)
\end{align}
Now for each $u \in B$ consider $\sum_{\tau \ge 0} \chi_{\tau,u} c_{k,\tau}(u)$.
%Clearly $\E[\sum_{\tau \ge 0} \chi_{\tau,u} c_{k,\tau}(u)] = \frac{1}{n} c_{k,0} + \E[\sum_{\tau \ge 1} \chi_{\tau,u} c_{k,\tau}(u)]$.
Is it immediate to see that $\E[\chi_{0,u}c_{k,0}] = \frac{1}{n} c_{k,0}$ and $\E[\sum_{\tau \ge 1} \chi_{\tau,u} c_{k,\tau}(u)] = \E[\sum_{\tau \ge 1} \big(\frac{1}{\ell} \sum_{j=1}^\ell \chi_{\tau,u}^j\big) c_{k,\tau}]$.
%Now consider the estimator $q_k^\ell(v)$ obtained by averaging $q_k(v)$ over the samples $\chi_{u,\tau}^1,\ldots,\chi_{u,\tau}^\ell$.
Therefore we replace $\sum_{\tau \ge 0} \chi_{\tau,u} c_{k,\tau}(u)$ with $\frac{1}{n} c_{k,0} + \sum_{\tau \ge 1} \big(\frac{1}{\ell} \sum_{j=1}^\ell \chi_{\tau,u}^j\big) c_{k,\tau}$.
It is clear that, if the latter summation is within a multiplicative $(1\pm\epsilon)$ of its own expectation, then the whole expression is as well.
%We want to show that, if point (2) of Lemma~\ref{lem:vplus2} holds, the last term of Equation~\ref{eqn:p_m_new_3} is within a multiplicative $(1\pm\epsilon)$ of its own expectation, as we did for PageRank.
By Equation~\ref{eqn:p_m_2_2}, $\sum_{\tau \ge 1} \big(\frac{1}{\ell} \sum_{j=1}^\ell \chi_{\tau,u}^j\big) c_{k,\tau}$ is a linear combination of $\sum_{\tau \ge 1} \big(\frac{1}{\ell} \sum_{j=1}^\ell \chi_{\tau,u}^j\big) c_{H,\tau}(u)$ over $H=G_{j(u)},\ldots,G_k$.
However $\sum_{\tau \ge 1} \big(\frac{1}{\ell} \sum_{j=1}^\ell \chi_{\tau,u}^j\big) c_{H,\tau}(u)$ equals $s_{H}(u) \, c_{H,1}$, which by Lemma~\ref{lem:vplus2} with probability $1-\frac{\delta}{2}$ is within a multiplicative $(1\pm\epsilon)$ of its own expectation for all $u \in B$.
We thus get a multiplicative $(1\pm\epsilon)$-approximation of the rightmost summation in Equation~\ref{eqn:p_m_new_3}, while for the remaining terms we use the concentration bounds of Subsection~\ref{sub:conc2}, as for PageRank.

\subsubsection{Indegree inequalities}
We adapt Section~\ref{sub:indegineq}.
First, we prove:
\begin{lemma}
\label{lem:pc2}
For any $u \in G$ it holds: \emph{1.} $\indegree{u} = O(\dmax n P_1(u) )$, \emph{2.} $\indegree{u}  = O((m n \ln(n) P_1(u))^{1/2})$.
\end{lemma}
\begin{proof}
Recall that $P_1(u) = \alpha \sum_{w \rightarrow u} \frac{P_0(w)}{\outdegree{w}}$ and $P_0(w) = \frac{\eta}{n}$.
Then use the proof of Lemma~\ref{lem:pc} (Appendix~\ref{apx:indegineq}).
\end{proof}
\noindent As a consequence, for the query complexity $t=\sum_{i=0}^k(1+2\,\indegree{u_i})$ of building $q_k(v)$ we get the same bounds of Equation~\ref{eqn:bound_t}.
Since the blacklisting and sampling phase have query complexity $O(\ell \ln^2(n/\epsilon))$ (see Subsection~\ref{sub:rwalk2}), we set $t = \ell \ln^2(n/\epsilon)$ to minimize the total query complexity.
As for PageRank we shall then set $k = \frac{m \ln(1/\delta)}{\epsilon^2 d \ell}$ (see Section~\ref{sub:conc2}).
After a few manipulations we get:
\begin{align}
\label{eqn:qc_3}
\ell = O\big(\!\min\!\big(m^{2/3} \dmax^{1/3} d^{-2/3} \ln(n/\delta)^{1/3} \ln(n/\epsilon)^{-2/3} \ln(1/\delta)^{1/3} \epsilon^{-4/3} ,\;\\
m^{4/5} d^{-3/5} \ln(n/\delta)^{1/5} \ln(n/\epsilon)^{-4/5} \ln(1/\delta)^{2/5}  \epsilon^{-6/5}
\big)\!\big)
\end{align}
which is $\tilde{O}\big(\!\min\!\big( m^{2/3} \dmax^{1/3} d^{-2/3}, m^{4/5} d^{-3/5}\big)\big)$ if we hide factors depending only on $\epsilon$ and $\delta$.

\subsubsection{Approximate estimators}
We just need to prove:
\begin{lemma}
\label{lem:pmv_approx_2}
An additive $\bareps$-approximation $p_k'(v)$ of $p_k(v)$ can be built in time $O(k t \ln(1/\bareps) \ln(n/\epsilon))$.
\end{lemma}
\begin{proof}
The proof is essentially the same as the proof of Lemma~\ref{lem:pmv_approx}  (Appendix~\ref{apx:pmv_approx}); the only difference is that we now have $n a \ln(n/\epsilon)$ coefficients rather than just $n$.
Therefore we need a $p_k'(v)$ whose generic coefficient differs additively by at most $\frac{\bareps}{2 a \ln(n/\epsilon)}$ from its counterpart in $p_k(v)$.
Suppose then $|c_k-c_k'| \le \frac{\bareps}{2 a \ln(n/\epsilon)}$ and $|c_k(u)-c_k'(u)| \le \frac{\bareps}{2 a \ln(n/\epsilon)}$ for all $u \in G$.
Then $|\E[p_k(v) - p_k'(v)]| \le |c_k-c_k'|+ \sum_{\tau \ge 0} \sum_{u \in G} |c_k(u)-c_k'(u)| \E[\chi_{\tau,u}] < \frac{\bareps}{2} + \sum_{\tau \ge 0} \frac{\bareps}{2 a \ln(n/\epsilon)} \sum_{u \in G} \E[\chi_u] = \bareps$, using the fact that $\sum_{u \in G}\E[\chi_u] = 1$ and that we consider only $a \ln(n/\epsilon)$ terms over $\tau \ge 0$.

The rest of the proof shall be adapted as follows.
We need to compute additive $\frac{\bareps}{2 a \ln(n/\epsilon)}$-approximations of the coefficients $c_{G_k}$ and $c_{G_k,\tau}(u)$ for all $\tau \ge 0$ and $u \in F(G_k)$; see Subsection~\ref{sub:subgraph2}.
For any $\tau$ and any $\lambda \ge 0$ then let:
\begin{align}
r_k^{\tau,\lambda}(u_i) = \frac{\alpha^{\tau+\lambda+1}}{(\tau+\lambda+1)!} \sum_{\substack{\pi \in \Pi_{G_k}(u_i,v) \\ |\pi|=\lambda}} \,\, \prod_{(w,w')\in \pi}\frac{1}{\outdegree{w}}
\end{align}
From the definitions in Subsection~\ref{sub:subgraph2}, then, $c_{G_k,\tau}(u) = \eta \frac{1}{\outdegree{u}} \sum_{w:(u,w)\in F(G_k)} \sum_{\lambda \ge 0} r_k^{\tau,\lambda}(u_i)$ and $c_{G_k} = \frac{\eta }{n} \sum_{z \in G_k} \sum_{\lambda \ge 0} r_k^{-1,\lambda}(z)$.
In other words $c_{G_k,\tau}(u)$ and $c_{G_k}$ are bounded by weighted averages of the $\sum_{\lambda \ge 0}r_k^{\tau,\lambda}(u_i)$ and therefore we just need a $\frac{\bareps}{2 a \ln(n/\epsilon)}$-approximation of this quantity.
For all $\lambda \ge 0$ let then $\mathbf{r}_k^{\tau,\lambda}$ be the vector whose $i$-th component is $r_k^{\tau,\lambda}(u_i)$, for $i=0,\ldots,k$.
For $\lambda=0$ we set $r_k^{\tau,0}(v) = 1$ and $r_k^{\tau,0}(u_i)=0$ for all $u_i \ne v$.
Let $\bfA_k \in \mathbb{R}^{(k+1)\times(k+1)}$ be the normalized adjacency matrix of $G$ restricted to $G_k$, so $\bfA_k[i,j] = \frac{1}{\outdegree{u_i}}$ if and only if $u_i,u_j \in G_k$ and $(u_i,u_j) \in G$.
Then for all $\lambda \ge 1$:
\begin{align}
\label{eqn:paths_matrix_series_2}
\mathbf{r}_k^{\tau,\lambda} = \frac{\alpha}{\tau+\lambda+1} \bfA_k \, \mathbf{r}_k^{\tau,\lambda-1}
\end{align}
Thereafter, the arguments of the proof of Lemma~\ref{lem:pmv_approx} (Appendix~\ref{apx:pmv_approx}) hold unchanged.
\end{proof}

As for PageRank, we can now pick $\bareps = \Theta(\frac{\epsilon}{n})$ and build $p_k'(v)$ in time $O(k t \ln^2(n/\epsilon))$.
The overall computational complexity is therefore the same of PageRank, save for polylogarithmic factors: $\tilde{O}\big(\min\!\big(m^{3/4} \dmax^{1/4} d^{-3/4}, m^{6/7} d^{-5/7} \big)\big)$.

\subsection{Proof of Theorem~\ref{thm:simple_ub}}
\label{apx:simple_ub}
\subsubsection{Upper bound}
We port our algorithm in the model of~\cite{Brautbar&2010}.
We show the adaptation only for PageRank, but it is straightforward to obtain the bounds for heat kernel after adapting the original algorithm as specified in Appendix~\ref{apx:me}.
First we show that, if the query \neigh[u] returned not only the parents and the children of $u$ but also the outdegree of each parent of $u$, then we would need only $O(n^{1/2})$ queries.
Then we show how, by sketching an approximation of the outdegrees, one need $\tilde{O}(n^{2/3})$ queries.
Note that we use our first estimator $p_k(v)$ (see Section~\ref{sub:exploring_choosing}), i.e.\ we do not need the blacklisting phase.

Suppose then \neigh[u] returns, in addition to the parents and the children of $u$, the outdegree of each parent of $u$.
Then invoking \neigh[u] suffices to expand $u$, hence we can build $p_k(v)$ with $O(k)$ queries (see Section~\ref{sub:exploring_choosing}).
Since the sampling phase takes $O(\ell)$ queries, then, we can minimise the query complexity by setting $\ell = k$.
By the bounds of Section~\ref{sub:conc} we can pick $k \ell = O(\frac{n\ln(1/\delta)}{\epsilon^2})$, which gives $k = \ell = O(n^{1/2}\ln(1/\delta)^{1/2} \epsilon^{-1})$.

Now suppose instead \neigh[u] returns an approximation of the outdegrees of $u$'s parents.
For each node $u \in G$ we denote by $\outestim{u}$ such an approximation.
For any pair of nodes $u,u' \in G$ with positive outdegrees let $r(u,u') = \frac{\outdegree{u}}{\outdegree{u'}}/\frac{\outestim{u}}{\outestim{u'}} \ge 1$.
Without loss of generality assume $r(u,u') \ge 1$ (otherwise just switch $u$ and $u'$).
In other words $r(u,u')$ tells by how much the ratio $\frac{\outdegree{u}}{\outdegree{u'}}$ changes if we use the outdegree approximations.
Now let:
\begin{align}
\gamma = \max\{ r(u,u'),\; u,u' \in G \,:\, \outdegree{u},\, \outdegree{u'} > 0\}
\end{align}
We then build $p_k(v)$ as usual (Section~\ref{sub:exploring_choosing}), but by using the approximations $\outestim{u}$ in place of the actual outdegrees $\outdegree{u}$ (the outdegrees appear at the denominator of the coefficients $c_H$ and $c_H(u)$ -- see Appendix~\ref{apx:PR_conductance}).
This means we choose $\beta_0,\ldots,\beta_k$ and $u_0,\ldots,u_k$ so that $p_k(v)$ is a perfect weighted estimator (Definition~\ref{def:balanced_estimator}).
Note however that $\E[p_k(v)] \ne P(v)$ in general.

Once built $p_k(v)$, we change all its coefficients by replacing $\outestim{u}$ with $\outdegree{u}$, which makes $\E[p_k(v)] = P(v)$.
This replacement can be performed directly for every $u \in G_k$ since we have queried it.
For the nodes $u \in F(G_k)$ instead we do not know $\outdegree{u}$.
However, to take a sample of $p_k(v)$ we only need to know the outdegree of the coefficient $c_k(u)$ associated to the node $u$ returned by \samplenode\ i.e.\ such that $\chi_u=1$, which we can learn with a single additional query.
All other terms are implicitly set to $0$ and therefore knowing their coefficients is irrelevant.

Now, by definition of  $c_k$, $c_k(u_i)$, $c_k(u)$ (Section~\ref{sub:exploring_choosing}) and in turn of $c_H$ and $c_H(u)$ (Appendix~\ref{apx:PR_conductance}) it is straightforward to see that the ratio of any two coefficients of $p_k(v)$ changes by no more than $\gamma$.
Formally $p_k(v)$ is $\gamma$-\emph{perfect}, where:
\begin{definition}
\label{def:gamma_balanced_estimator}
We say the weighted estimator $p_k(v)$ is $\gamma$-\emph{perfect} if:
\begin{align*}
\begin{array}{ll}
c_k(u_i) \ge \frac{1}{\gamma} c_k(u_{i'}) &\text{for all } i,i' \in \{1,\ldots,k\}\\[5pt]
c_k(u_i) \ge \frac{1}{\gamma} c_k(u) &\text{for all } i \in \{1,\ldots,k\} \text{ and all } u \in F(G_k)
\end{array}
\end{align*}
\end{definition}
We now adapt the concentration bounds of Section~\ref{sub:conc}.
Let $c = \max\{ \{c_k(u_i) : i=1,\ldots,k \} \cup \{c_k(u) : u \in F(G_k)\}\}$.
Then the random variable $c^{-1}(p_k(v) - c_k)$ is a sum of non-positively correlated random variables with coefficients in $[0,1]$ and Equation~\ref{eqn:pml_bound_2} holds.
Now:
\begin{align}
\E[c^{-1}(p_k(v) - c_k)] \ge \E\Big[c^{-1} \sum_{i = 1}^{k} \chi_{u_i} c_k(u_i)\Big] \ge \frac{1}{\gamma} \E\Big[\sum_{i = 1}^{k} \chi_{u_i}\Big]
\end{align}
where we used Definition~\ref{def:gamma_balanced_estimator} to bound $c_k(u_i) \ge c \frac{1}{\gamma}$.
Finally, since $\E[\chi_{u_i}] = P(u_i) \geq \frac{1-\alpha}{n}$, we obtain $\E[c^{-1}(p_k(v) - c_k)] \ge \frac{1-\alpha}{n\gamma}k$, and if we take $\ell$ independent samples of $p_k(v)$ the expectation grows to $\frac{1-\alpha}{n\gamma}k\ell$.
Hence, as in Equation~\ref{eqn:pml_bound_2}, we obtain:
\begin{align}
\label{eqn:conc_pkl_gamma}
\prob\big[\,|p_k^{\ell}(v) - P(v)| > \epsilon P(v)\big] \leq 2\exp\!\Big(\!-\frac{\epsilon^2 (1-\alpha) k \ell }{3\gamma n}\Big)
\end{align}

To get our multiplicative $(1\pm\epsilon)$-approximation of $P(v)$ with probability $1-\delta$ we must then pick $k \ell = \Theta(n \gamma \epsilon^{-2} \ln(1/\delta))$.
We shall now show that, by spending $k$ queries to approximate the outdegrees, one can essentially guarantee $\gamma \le \frac{n \ln(n)}{k}$.
Replacing this bound at the exponent of Equation~\ref{eqn:conc_pkl_gamma} and optimizing for $k$ will prove our upper bound.

\begin{lemma}
\label{lem:approx_degrees}
With $k$ \jump{} and $k$ \neigh[\cdot] queries one can obtain estimates $\{\outestim{u}\}_{u \in G}$ such that, for any $b > \frac{2k}{n\ln(n)}$, we have $\gamma \leq 4b \frac{n \log(n)}{k}$ with probability $1 - 2n^{-\frac{b}{8 \ln(2)}+1}$.
\end{lemma}
\begin{proof}
We draw $k$ nodes from $G$ using \jump{}, ad one each of them we invoke \neigh\ to learn its parents.
For each $u\in G$ let then $\psi_u^j$ be the indicator random variable of the event that $u$ is a parent of the $j$-th node drawn; clearly $\E[\psi_u^j] = \frac{\outdegree{u}}{n}$.
For every $u \in G$ and all $b>0$ let:
\begin{align}
\outestim{u} = \frac{n}{k}\Big(\sum_{j=1}^k \psi_u^j + b \log(n)\Big)
\end{align}
Note that $\E[\outestim{u}] = \outdegree{u} + \frac{n}{k}b \log(n)$ and that $\sum_{j=1}^k \psi_u^j$ is a sum of independent binary random variables.
We now prove that, with probability $1 - 2n^{-\frac{b}{8 \ln(2)}+1}$, for all $u$ with $\outdegree{u} > 0$ it holds $\frac{1}{2} \le \frac{\outestim{u}}{\outdegree{u}} \le 2 \frac{n}{k} b \log(n)$.
By definition of $\gamma$ this proves the theorem.

Let us start with the lower bound.
Suppose $\outdegree{u} \leq \frac{n}{k}b\log(n)$; this implies $\frac{\outestim{u}}{\outdegree{u}} \ge 1 > \frac{1}{2}$.
Suppose instead $\outdegree{u} > \frac{n}{k}b\log(n)$.
Then $\E[\sum_{j=1}^{k} \psi^j_{u}] > b\log(n)$, and by the probability bounds of Appendix~\ref{apx:chernoff_bounds} we have $\prob\big[\sum_{j=1}^{k} \psi^j_{u} < \frac{1}{2}\E[\sum_{j=1}^{k} \psi^j_{u}]\big] < e^{-\frac{b\log(n)}{8}} = n^{-\frac{b}{8 \ln(2)}}$.
This is also a bound on $\prob[\outestim{u} < \frac{1}{2}\E[\outestim{u}]]$ by construction of $\outestim{u}$, and since $\outdegree{u} < \E[\outestim{u}]$, on the probability that $\outestim{u} < \frac{\outdegree{u}}{2}$.
Taking a union bound on all $u$, the probability that $\frac{\outestim{u}}{\outdegree{u}} < \frac{1}{2}$ for some $u$ is at most $n^{-\frac{b}{8 \ln(2)}+1}$.

Let us now turn to the upper bounds.
Note that $\frac{n}{k} \sum_{j=1}^k \psi_u^j  + \frac{n}{k} b \log(n) \outdegree{u} \ge \outestim{u}$ since $\outdegree{u} \ge 1$.
Hence the event $\frac{\outestim{u}}{\outdegree{u}} > 2 \frac{n}{k} b \log(n)$, or equivalently $\outestim{u} > 2 \frac{n}{k} b \log(n) \outdegree{u}$, implies $\sum_{j=1}^k\!\psi_u^j > b \log(n) \outdegree{u}$.
However, $\E[\sum_{j=1}^k\!\psi_u^j] = \frac{k}{n}\outdegree{u}$.
The event is thus equivalent to $\sum_{j=1}^k\!\psi_u^j = \E[\sum_{j=1}^k\!\psi_u^j](1+\epsilon)$ for some $\epsilon \ge \frac{n b \ln(n)}{k}-1$.
The probability of this event is, by the bounds of Appendix~\ref{apx:chernoff_bounds}, smaller than $e^{-\frac{\epsilon^2}{2+\epsilon}\E[\sum_{j=1}^k\!\psi_u^j]}$.
Note however that since $b > \frac{2k}{n\ln(n)}$ then $\epsilon > 1$, which implies $\frac{\epsilon^2}{2+\epsilon} > \frac{1+\epsilon}{6}$.
At the exponent of the bounds we can then just plug $\frac{b \log(n) \outdegree{u}}{6}$ in place of $\frac{\epsilon^2}{2+\epsilon} \E[\sum_{j=1}^k \psi_u^j]$, obtaining an upper bound of $e^{-\frac{b\log(n)\cdot \outdegree{u}}{6}} \leq n^{-\frac{b}{6\ln(2)}}$.
Taking a union bound on all $u$, the probability that $\frac{\outestim{u}}{\outdegree{u}} > 2 \frac{n}{k} b \log(n)$ for some $u$ is at most $n^{-\frac{b}{6\ln(2)}+1}$.

A final union bound completes the proof.
\end{proof}

We can now conclude the proof of the upper bound in Theorem~\ref{thm:simple_ub}.
First, we use the same number of queries $k$ for approximating the outdegree, building $p_k(v)$, and sampling $p_k(v)$.
By Lemma~\ref{lem:approx_degrees}, we can make arbitrarily smaller than $\delta$ the probability that $\gamma > 4b\frac{n \log(n)}{k}$ by choosing a large enough $b \in O(1)$.
We can then assume $\gamma \leq 4b\frac{n \log(n)}{k}$.
Then by Equation~\ref{eqn:conc_pkl_gamma}:
\begin{align}
\label{eqn:pml_bound_3}
\prob\big[\,|p_k^{\ell}(v) - P(v)| > \epsilon P(v)\big] \leq 2\exp\!\Big(\!-\frac{\epsilon^2 (1-\alpha) k^3 }{12 \,b\, n^2 \ln(n) }\Big)
%\leq 2e^{-(\frac{\epsilon^2}{3} \frac{k^3}{4 b n \log(n)} \frac{1-\alpha}{n})}
\end{align}
To make the right-hand side arbitrarily smaller than $\delta$, it suffices to pick:
\begin{align}
k = O\big(n^{\frac{2}{3}}\ln(n)^{1/3}\,\ln(1/\delta)^\frac{1}{3}\epsilon^{-\frac{2}{3}}\big) = \tilde{O}\big(n^{\frac{2}{3}}\big)
\end{align}
concluding our query complexity upper bound.

\subsubsection{Lower bound}
It suffices to adapt the graph for the lower bounds of~\ref{sec:lowerbounds}.
Set the number of parents of $v$ to $g=\Theta(n^{2/3})$, and the number of children of each parent to $\Delta=\Theta(n^{1/3})$.
Crucially, distinct parents must have no child in common.
Set $\gamma=\Theta(\Delta)$.
It is immediate to check that $P(v) = \Theta(n^{-2/3})$ and that we can change $P(v)$ by constant factors by switching the rightmost parent between having either $\gamma$ parents or $\gamma$ children.
Clearly, one cannot distinguish between the two cases unless one queries the rightmost parent or one of its $\gamma$ neighbors.
To simplify the analysis we can reinforce the model by assuming that (1) \neigh[v] is given for free at the beginning, (2) any \jump{} returns a node $u$ as well as its parents, its children, siblings (children of its parents), and spouses (parents of its children).
Now, for any node $u \ne v$ returned by \jump{} or queried by the algorithm via \neigh[u], we can mark $u$ and all of its parents/children/siblings/spouses as visited.
There are therefore $n^{2/3}$ subsets of nodes to be marked as visited, and it is easy to see that finding the rightmost parent of $v$ among them takes in expectation either $\Omega(\frac{n}{\gamma}) = \Omega(n^{2/3})$ queries via \jump{} or $\Omega(\Delta) = \Omega(n^{2/3})$ queries via \neigh.